\documentclass{aastex631}

\def\cof {$^{12}\mathrm{CO}~(J=1\rightarrow0)$}
\def\cos {$^{13}\mathrm{CO}~(J=1\rightarrow0)$}

\def\wfour{G134.8$+$01.4}
\def\largestMC{G109.7$+$01.8}
\def\cot {$\mathrm{C}^{18}\mathrm{O}~(J=1\rightarrow0)$}
\def\msun{$M_{\odot}$}
 
\def\ag{$A_{G}$}
\def\av{$A_{V}$}
\def\cofs {$^{12}\mathrm{CO}$}

\def\deg  {\ifmmode {^\circ}\else {$^\circ$}\fi}

\def\kms     {km~s$^{-1}$}

\usepackage{booktabs}
\usepackage{chngcntr}

\date{\today}
\submitjournal{ApJ}

\shorttitle{Improved Distance Measurements}
\shortauthors{Yan et al.} 

\begin{document}


\title{Improved Measurements of Molecular Cloud Distances Based on Global Search}

\correspondingauthor{Ji Yang}
\email{jiyang@pmo.ac.cn,qzyan@pmo.ac.cn}

\author[0000-0003-4586-7751]{Qing-Zeng Yan}
\altaffiliation{Chang Yu-Che Fellow of Purple Mountain Observatory}
\affil{Purple Mountain Observatory and Key Laboratory of Radio Astronomy,\\
 Chinese Academy of Sciences, 10 Yuanhua Road, Qixia District, Nanjing 210033, People's Republic of China}

\author[0000-0001-7768-7320]{Ji Yang}
\affil{Purple Mountain Observatory and Key Laboratory of Radio Astronomy,\\
 Chinese Academy of Sciences, 10 Yuanhua Road, Qixia District, Nanjing 210033, People's Republic of China}

 \author[0000-0002-0197-470X]{Yang Su }
 \affil{Purple Mountain Observatory and Key Laboratory of Radio Astronomy,\\
 Chinese Academy of Sciences, 10 Yuanhua Road, Qixia District, Nanjing 210033, People's Republic of China}
 
 \author[0000-0002-3904-1622]{Yan Sun}
\affil{Purple Mountain Observatory and Key Laboratory of Radio Astronomy,\\
 Chinese Academy of Sciences, 10 Yuanhua Road, Qixia District, Nanjing 210033, People's Republic of China}
 
  \author[0000-0001-5602-3306]{Ye Xu}
\affil{Purple Mountain Observatory and Key Laboratory of Radio Astronomy,\\
 Chinese Academy of Sciences, 10 Yuanhua Road, Qixia District, Nanjing 210033, People's Republic of China}
 
 \author[0000-0003-0746-7968]{Hongchi Wang}
\affil{Purple Mountain Observatory and Key Laboratory of Radio Astronomy,\\
 Chinese Academy of Sciences, 10 Yuanhua Road, Qixia District, Nanjing 210033, People's Republic of China}
 
 \author[ 0000-0003-2418-3350]{Xin Zhou}
\affil{Purple Mountain Observatory and Key Laboratory of Radio Astronomy,\\
 Chinese Academy of Sciences, 10 Yuanhua Road, Qixia District, Nanjing 210033, People's Republic of China}

  \author[0000-0001-8923-7757]{Chen Wang}
 \affil{Purple Mountain Observatory and Key Laboratory of Radio Astronomy,\\
 Chinese Academy of Sciences, 10 Yuanhua Road, Qixia District, Nanjing 210033, People's Republic of China}





\begin{abstract} 
The principle of the background-eliminated extinction-parallax (BEEP) method is examining the extinction difference between on- and off-cloud regions to reveal the extinction jump caused by molecular clouds, thereby revealing the distance in complex dust environments. The BEEP method requires high-quality images of molecular clouds and high-precision stellar parallaxes and extinction data, which can be provided by the Milky Way Imaging Scroll Painting (MWISP) CO survey and the Gaia DR2 catalog, as well as supplementary \av\ extinction data. In this work, the BEEP method is further improved (BEEP-II) to measure  molecular cloud distances in a global search manner. Applying the BEEP-II method to three regions mapped by the MWISP CO survey, we collectively measured  238 distances for 234 molecular clouds. Compared with previous BEEP results, the BEEP-II method measures distances efficiently, particularly for those molecular clouds with large angular size or in complicated environments, making it suitable for distance measurements of molecular clouds in large samples.
\end{abstract}




\section{Introduction} \label{sec:intro}

The distance is a fundamental property of molecular clouds and is usually an essential part of molecular cloud catalogs \citep[e.g.,][]{2017ApJ...834...57M,2019MNRAS.483.4291C}. Although distances are not intrinsic, they are necessary for deriving physical properties of molecular clouds, such as the mass and size. 




Molecular cloud catalogs were empirically built with spectroscopic data in position-position-velocity (PPV) space. For instance, \citet{1979IAUS...84...35S} and \citet{1987ApJ...319..730S} used closed contours to define Giant Molecular Clouds (GMCs), and apparently, different contour levels correspond to different molecular cloud  samples. With the expansion of data, mathematical algorithms were involved to identify molecular clouds automatically in PPV space. For example,  \citet{2016ApJ...822...52R} identified 1064 high-mass molecular clouds using dendrograms and the CfA-Chile survey \citep{2001ApJ...547..792D}, while with the same data but an alternative hierarchical clustering algorithm, \citep{2017ApJ...834...57M} found 8107 molecular clouds.

 However, in some regions, where molecular clouds are crowded in PPV space, even mathematical algorithms are incapable of identifying molecular clouds properly. In addition to the data quality effects, such as the resolution and sensitivity \citep{2021ApJ...910..109Y}, the information provided by PPV data is intrinsically incomplete. PPV data miss one dimension of position, and the relationship between the radial velocity and the distance is not always one-to-one. Here, we list three situations when identifying molecular clouds can be problematic. First, low angular resolution observations would artificially merge those molecular clouds that are close to each other (e.g., $<$ 1 pc), and high angular resolutions are needed to distinguish them. Secondly, in the first and fourth Galactic quadrants, one radial velocity may correspond to both near and far distance components \citep{2009ApJ...700..137R}, and due to the cloud-cloud velocity dispersion \citep[about 5 \kms,][]{2021ApJ...910..131S}, the association between the radial velocity and the distance of local molecular clouds is also weak. Thirdly, toward the Galactic anticenter region (e.g., $l=180^\circ$), the radial velocity is close to zero and provides little information about the distance. Consequently, measuring distances to molecular clouds identified in PPV space is particularly important.

Various methods have been developed to obtain molecular cloud distances, but they all have limitations. Those methods can be broadly categorized as either ``low accuracy" or ``high accuracy". For instance, counting stars \citep{1923AN....219..109W} is a low accuracy method that suffers from the large systematic errors introduced by the assumed distribution of stars. Another low accuracy method is to consult close (judged by the angular distance) OB-associations \citep{2019MNRAS.487.2522M}, but the relationships between OB-associations and their nearby molecular clouds are largely unclear. The widely-used kinematic distances, which are based on the rotation curve of the Milky Way, also yield large errors, particularly for local molecular clouds \citep[e.g.,][]{2009ApJ...700..137R}.

Measuring molecular cloud distances accurately is difficult, but achievable. For example, one can use the trigonometric parallaxes of masers as proxies for distances of their host molecular clouds \citep{2006Sci...311...54X,2019ApJ...885..131R,2020PASJ...72...50V}, and this maser-parallax method is particularly useful for high-mass star-forming regions. However, not all molecular clouds have maser emission \citep{2021MNRAS.500.3027D}, and being point sources, masers may only poorly represent distances of extended structures \citep{2018ApJ...869...83Z}. Young stellar objects (YSOs) is also able to reveal the distances of their natal molecular clouds, and in addition, YSOs can be used to investigate the structure and dynamic properties of local star-forming regions \citep[e.g.,][]{2020A&A...638A..85R}. This YSO method is only applicable to nearby star-forming regions \citep{2018A&A...620A.172Z,2018ApJ...867..151D}, because distant YSOs are faint in the visible band due to the interstellar extinction. The interstellar extinction, however, is not always a bad thing, as it provides another approach to infer molecular cloud distances. In principle, the extinction of stars behind (background) molecular clouds is larger than that of stars in front of (foreground) molecular clouds along the line of sight, and the extinction jump at specific distances corresponds to molecular cloud distances. This extinction method belonged to low accuracy methods, but its accuracy has been improved significantly owing to the release of the Gaia DR2 catalog \citep{2016A&A...595A...1G,2018A&A...616A...1G}.

  



In regions where molecular clouds are well-separated in both position-position-position (PPP) and PPV space, the difference between foreground and background stellar extinction is sufficient to reveal molecular cloud distances. For example, \citet{2019ApJ...879..125Z,2020A&A...633A..51Z} and \citet{2019A&A...624A...6Y} obtained distances to many molecular clouds at high Galactic latitudes, with distance errors usually less than 10\%.

However, large-scale CO surveys usually focus on the Galactic plane, where the dust environment is complicated. In addition, with the increase of sensitivity and angular resolutions, those surveys detect a large number of molecular clouds that have no distance indicators \citep[e.g.,][]{2021A&A...645A.129Y}, such as masers, YSOs, or adjacent OB stars. Consequently, we need more sophisticated methods to measure molecular cloud distances for large-scale sensitive CO surveys.




\citet{2019ApJ...885...19Y} proposed a background-eliminated extinction-parallax (BEEP) method to accurately measure molecular cloud distances in the Galactic plane, based on CO maps of the Milky Way Imaging Scroll Painting (MWISP) survey \citep{2019ApJS..240....9S} and the Gaia DR2 catalog. The BEEP method classifies on- and off-cloud stars with high-quality CO maps, and uses off-cloud stars to remove the extinction background of on-cloud stars, thereby revealing molecular distances accurately. The BEEP method has been shown to work well in three regions mapped by the MWISP survey \citep{2019ApJ...885...19Y,2020ApJ...898...80Y,2021A&A...645A.129Y}.

Nonetheless, applying the BEEP method to large or complicated molecular clouds is not straightforward. Large-scale CO surveys with high sensitivity identify many molecular clouds with large angular sizes or in complicated environments \citep[e.g., see the ``Phoenix" and ``River" clouds of][]{2020ApJ...893...91S}, causing two problems in applying the BEEP method: (1) low efficiency of selecting pairs of on- and off-cloud stars and (2) significant variation of extinction background across molecular clouds. The BEEP method requires pairs of on- and off-cloud stars, and the choice of on- and off-cloud star samples is mostly determined by eye. For example, the on- and off-cloud regions are usually chosen near the edge of molecular clouds where the gradient of the integrated intensity is large. Several pairs of on- and off-cloud stars may need to be tested before a reliable distance is derived. Evidently, this requires careful scrutiny of individual cases, is too laborious and time-consuming to be practical for large samples, and the yielding results may not be optimal.

In this work, we refine the BEEP method to make it more feasible. The improved method, BEEP-II, is designed to obtain comprehensive distances via building pairs of on- and off-cloud stars across molecular clouds and applying the BEEP method multiple times, so it should improve the distance precision and isolate those molecular clouds that have multiple distance components. The BEEP-II method is expected to be able to measure distances to  molecular clouds with large angular size or in complicated environments.

This paper is organized as follows. The next section (Section \ref{sec:data}) describes the CO and Gaia DR2 data, as well as the cloud identification method. Section \ref{sec:besd} demonstrates the procedure of the BEEP-II method. Section \ref{sec:result} displays distances of three regions mapped by the  MWISP survey. Discussions are presented in Section \ref{sec:discuss}, and we summarize the conclusions in Section \ref{sec:summary}.









\section{CO Data and Molecular Cloud catalogs} 
\label{sec:data}

{\catcode`\&=11
\gdef\2021AandA...645A.129Y{\citet{2021A&A...645A.129Y}}}
\begin{deluxetable}{cccccccc}
\tablecaption{Molecular Cloud Distances in the Literature Based on the MWISP CO Survey. \label{Tab:disMWISP}}
\tablehead{
\colhead{Quadrant} & \colhead{$l$} & \colhead{$b$}  &  \colhead{$V_{\rm LSR}$}   & \colhead{Algorithm} & \colhead{Total Number\tablenotemark{a} } & 
\colhead{Cloud Distances} & \colhead{Reference} \\
\colhead{   } & \colhead{(\deg)} & \colhead{(\deg)} & (km/s) &  &    &  
}

\startdata
First  &  [25.8, 49.7] &  [-5, 5]& [-6, -30]& SCIMES\tablenotemark{b} & 898   & 28 &  \citet{2020ApJ...898...80Y}  \\ 
Second  &  [104.75, 150.25] &  [-5.25, 5.25]& [-95, 25]& DBSCAN  & 1677   & 98\tablenotemark{c} &  \2021AandA...645A.129Y \\ 
Third  &  [209.75, 219.75] &  [-5, 5]& [0, 70]& dendrograms\tablenotemark{d} & 31   & 11 &  \citet{2019ApJ...885...19Y}  \\ 
\enddata
\tablenotetext{a}{This only takes into account molecular clouds with angular areas larger than 0.015 deg$^2$, including incomplete ones in PPV space.}
\tablenotetext{b}{Spectral Clustering for Interstellar Molecular Emission Segmentation \citep[SCIMES,][]{ 2015MNRAS.454.2067C}.}
\tablenotetext{c}{Including 22 distances of the largest molecular cloud, see Figure 5 of \2021AandA...645A.129Y.}
\tablenotetext{d}{See \citet{2008ApJ...679.1338R} for details of dendrograms.}
\end{deluxetable}

\subsection{CO and Gaia DR2 Catalog} 

The CO data are part of the MWISP\footnote{\href {http://www.radioast.nsdc.cn/mwisp.php}{http://www.radioast.nsdc.cn/mwisp.php}} CO survey \citep{2019ApJS..240....9S} conducted with the Purple Mountain Observatory (PMO) 13.7 m millimeter telescope. The MWISP survey maps three CO isotopologue lines, \cof, \cos, and \cot; however, only \cofs\ is used in this work. The angular and velocity resolutions of the \cofs\ maps are approximately 49\arcsec\ and 0.2 \kms, respectively. CO maps are regridded into pixels of 30\arcsec, corresponding to an rms noise of about 0.49 K.

In this work, the BEEP-II method is applied to three regions that have been mapped by the MWISP survey. Table \ref{Tab:disMWISP} lists their $l$, $b$, and $V_{\rm LSR}$ ranges, as well as the BEEP results \citep{2019ApJ...885...19Y,2020ApJ...898...80Y,2021A&A...645A.129Y}.


The Gaia DR2 catalog \citep{2016A&A...595A...1G,2018A&A...616A...1G} provides an enormous number of  measurements of stellar parallaxes and extinction in the $G$ band (\ag), as well as an extra $V$-band  extinction (\av) catalog \citep{2019A&A...628A..94A}. Although the Gaia Early Data Release 3 (EDR3) improves the parallax precision by 30\%  \citep{2020arXiv201201533G}, the \ag\ and \av\ data have not been updated, so we still use the Gaia DR2 data. Stellar parallax errors are required to be less than 20\%, and all stars farther than 4 kpc ($<$ 0.25 mas) are removed. Corrections of systematic parallax errors follow \citet{2019A&A...628A..94A}, and in the distance measurement, \ag\ and \av\ extinction data are used separately.

\subsection{Molecular Cloud Samples}
 \label{sec:cloudIdentification}
We use the DBSCAN\footnote{\href {https://scikit-learn.org/stable/modules/generated/sklearn.cluster.DBSCAN.html}{https://scikit-learn.org/stable/modules/generated/sklearn.cluster.DBSCAN.html}} algorithm \citep{2020ApJ...898...80Y} to draw molecular cloud samples from PPV data cubes. Integrated intensity maps of molecular clouds identified with DBSCAN usually have clear edges, suitable for distance measurements with the BEEP method, which requires off-cloud regions to remove unrelated extinction fluctuations.


DBSCAN determines regions of molecular clouds in PPV space with two parameters, eps and MinPts. Eps is a distance threshold in specified metrics, and if the distance between two points is less than or equal to eps, they are considered neighbors. For a three-dimensional array, eps has three canonical values, 1, $\sqrt{2}$, and $\sqrt{3}$, which are referred to as connectivity 1, 2 and 3, respectively. Once the connectivity is assigned, MinPts specifies the minimum neighbors for a voxel to be considered a core point. Connected core points, as well as their neighbors, define molecular clouds.

\citet{2020ApJ...898...80Y} found that low MinPts values readily identify noise fluctuations and high MinPts values cause high fractions of missed flux. They suggested intermediate MinPts values for each connectivity, and the difference between three DBSCAN parameter combinations is insignificant. In this work, we adopt the same parameter settings with \citet{2021A&A...645A.129Y}, i.e., MinPts = 4 and connectivity = 1, because connectivity 1 possesses the highest compactness, and the DBSCAN parameters are constant across the PPV space. The brightness cutoff of CO data cubes is 2$\sigma$, about 1 K for \cofs. $\sigma$ is calculated for each spectrum, approximated by the rms of negative values. Criteria used to filter noise structures are also applied: (1) the minimum voxel number is 16; (2) the minimum peak brightness temperature is 5$\sigma$; (3) the integrated map contains at least one compact 2$\times$2 (60\arcsec$\times$60\arcsec) region; (4) the minimum number of velocity channels is 3. The third criterion requires at least a complete beam in the integrated map.

In addition to the above-mentioned four criteria, we further remove molecular clouds that touch the edge of PPV data cubes. Only a fraction of the mass and angular areas is observed for those molecular clouds, and since the fraction is unknown, they are excluded in the distance examination.


Molecular samples in the second Galactic quadrant are identical with those of \citet{2021A&A...645A.129Y}, except for the removal of incomplete clouds, but they are significantly different from those in the first \citep{2019ApJ...885...19Y} and third \citep{2020ApJ...898...80Y} Galactic quadrant, due to the use of alternative algorithms.




 \section{The BEEP-II method} 
\label{sec:besd}

\begin{figure}[ht!]
 
 \plotone{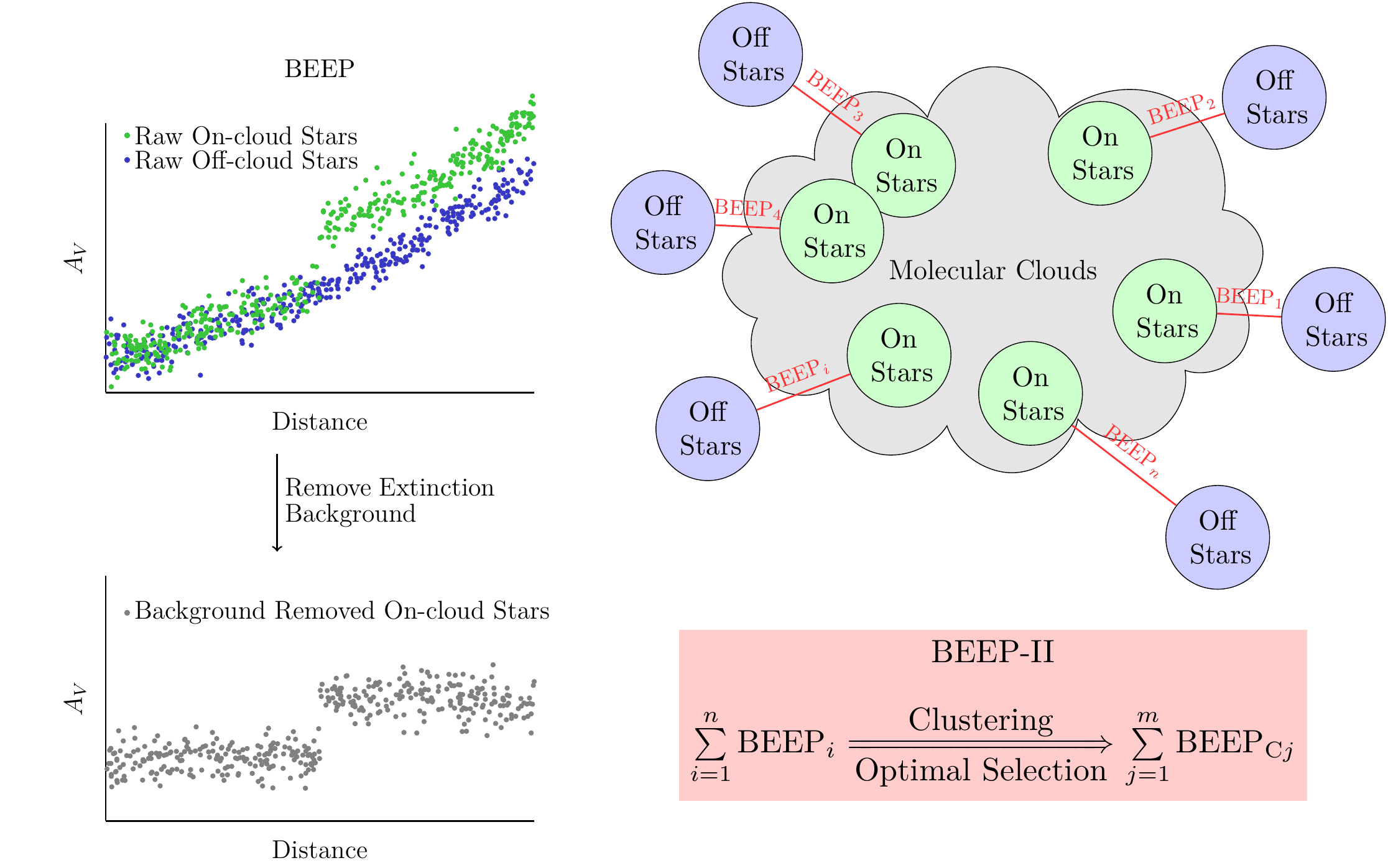}
 
\caption{Principles of the BEEP and BEEP-II methods. The green and blue dots represent variations of \av\ against the heliocentric distances for on- and off-cloud stars, respectively. The BEEP method removes extinction background in on-cloud stars with off-cloud stars, while the BEEP-II method examines multiple pairs of on- and off-cloud stars along the edge of molecular clouds. $\sum$ signifies combining multiple BEEP cases, and the collection of BEEP$_{\mathrm{C}j}$ is the optimal cases selected for clusters of BEEP$_i$. \label{fig:beepII} } 
\end{figure}

In this section, we present details of the BEEP-II method, an improved version of the BEEP method. A visualization of the BEEP and the BEEP-II methods is presented in Figure \ref{fig:beepII}.

The BEEP method \citep{2019ApJ...885...19Y} is developed to measure molecular cloud distances in the Galactic plane. Due to the dust component, background stars in on-cloud regions have, on average, larger extinction than those in off-cloud regions. For regions at high Galactic latitudes, there is usually only one nearby molecular cloud along each line of sight, and the variation of extinction is regular \citep{2019A&A...624A...6Y}. Specifically, the extinction of both foreground and background on-cloud stars shows Gaussian distributions. However, in the Galactic plane where molecular clouds are crowded, the extinction variations are irregular, causing gradients, and special treatments are needed. In principle, the BEEP method uses the extinction of off-cloud stars as a baseline, thereby removing the extinction variation that is unrelated to targeting molecular clouds.

 The BEEP method requires two data sets: high-quality images of molecular clouds and large numbers of stellar parallax and extinction measurements. Images of molecular clouds are used to distinguish on- and off-cloud stars, while stellar parallax and extinction are used to reveal the extinction jump caused by molecular clouds. It has been shown that the BEEP method is effective when applied to the MWISP CO survey and the Gaia DR2 catalog \citep{2019ApJ...885...19Y,2020ApJ...898...80Y,2021A&A...645A.129Y}. However, the BEEP method hinges on the selection of on- and off-cloud stars, which is usually performed with a trial-and-error strategy. The efficiency of this strategy is low, and it is hard to apply the BEEP method to molecular clouds with large angular size or in complicated environments. 


Here, we propose a refinement of the BEEP method, called BEEP-II, which is composed of three steps. First, we explore parameter combinations to select pairs of on- and off-cloud stars, deriving distances with the BEEP method for each pair. Secondly, because distances should be detected multiple times for a molecular cloud with a clear extinction jump, we cluster distances of a molecular cloud in 3D physical space and select the optimal parameter combination for each cluster. In the third step, we double check distances of clusters via MCMC sampling.


\subsection{Parameter Combinations}

To select a pair of on- and off-cloud stars, the BEEP-II method requires four parameters. Each pair of on- and off-cloud stars is confined within a circular region, which contributes two parameters: a center ($C_0$) and a radius ($R_0$). The remaining two parameters are the distance cutoff ($D_{\rm cut}$, in pc) and on-cloud CO cutoff ($\rm CO_{cut}$, in K \kms). All stars with distances larger than $D_{\rm cut}$ are removed, and the setting of $D_{\rm cut}$ avoids the interference of distant background stars, which is necessary for measuring distances of nearby molecular clouds accurately.

On-cloud $\rm CO_{cut}$ and off-cloud $\rm CO_{cut}$ have different meanings. On-cloud stars are removed if the integrated intensity toward them is less than on-cloud $\rm CO_{cut}$,  while off-cloud stars are removed if the integrated intensity toward them is larger than off-cloud $\rm CO_{cut}$. On-cloud $\rm CO_{cut}$ removes background stars that have low extinction, making the extinction jump clear, while off-cloud $\rm CO_{cut}$ removes stars that may have been significantly affected by other molecular clouds. High levels of on-cloud $\rm CO_{cut}$ can only be used when on-cloud stars are sufficient to trace foreground and background extinction distributions.

The $D_{\rm cut}$ list is 300, 500, 1000, 1500, 2000, 2500, and 3000 pc. As demonstrated by Figure 3 of \citet{2019A&A...624A...6Y}, the $D_{\rm cut}$ effect is insignificant as long as the extinction jump is clear. Ideally, a fixed $D_{\rm cut}$ of 3000 pc is sufficient, given that the extinction distributions of both foreground and background stars are all Gaussian. However, for nearby molecular clouds, we use large angular areas (usually larger than 1 square degree) to collect as many on-cloud stars as possible, and it is important to remember that large angular areas are likely to include other molecular clouds along the line of sight at far distances, deforming the Gaussian distribution of background stellar extinction. $D_{\rm cut}$ removes distant stars to make sure that background stars are not too far from each other. Since we have no prior information about molecular cloud distances, we try many $D_{\rm cut}$ values. Considering the distance errors and the insignificance of $D_{\rm cut}$ effects, a step of 500 pc suffices.

The on-cloud $\rm CO_{cut}$ list is 2, 3, 4, and 5 K \kms, while the off-cloud $\rm CO_{cut}$ is 1 K \kms, which is fixed. Those settings are determined according to the rms noise of the integrated intensity maps, which is about 1 K \kms. The off-cloud $\rm CO_{cut}$ corresponds to about 1$\sigma$, and the on-cloud $\rm CO_{cut}$ ranges from 2$\sigma$ to 5$\sigma$. On-cloud $\rm CO_{cut}$ removes low-extinction stars to increase the contrast between background and foreground extinction, at the cost of decreasing the number of on-cloud stars. Consequently, there is a trade-off between the clearness of extinction jump and the number of on-cloud stars. \citet{2021A&A...645A.129Y} use a fixed value of 4 K \kms\ for on-cloud $\rm CO_{cut}$, but for small molecular clouds, lower $\rm CO_{cut}$ values are better because  lower on-cloud $\rm CO_{cut}$ involves more on-cloud stars and yields more accurate distance results. There is also a trade-off for off-cloud $\rm CO_{cut}$, with 1$\sigma$ threshold retaining most off-clouds stars that are not significantly affected by other molecular clouds.

 $C_0$ and $R_0$  determine the circle region that contains on- and off-cloud stars. Considering $R_0$ as an angular size, it will be generally larger for nearby molecular clouds than for distant molecular clouds of comparable physical size. On the other hand, if we consider $R_0$ to represent physical size, then it is independent of distance. However, the conversion between physical and angular sizes requires prior knowledge of the distance, which may be estimated to be $D_{\rm cut}/2$.

The $C_0$ list is built from edge points (pixels) of integrated intensity maps of molecular clouds. (Voxels outside cloud regions are masked before the integration.) Examining all edge points is computationally expensive, so we designed a way to select a subset of edge points that are well separated from each other but are still dense enough to discern molecular cloud components. The typical size of molecular clouds is 10 pc \citep{2015ARA&A..53..583H}, so we use 5 pc as a separation between $C_0$. The angular size that corresponds to 5 pc is calculated by assuming that molecular clouds are at a distance of $D_{\rm cut}/2$. Initially, we measure all angular distances from edge points to the centroid position of molecular clouds, and append two edge points, corresponding to the largest and smallest angular distance, to the $C_0$ list. The centroid position, provided by the molecular cloud catalog \citep{2020ApJ...898...80Y}, of molecular clouds is the average $l$ and $b$ (of voxels) weighted by the brightness temperature. Secondly, we remove edge points that are within 5 pc of the current $C_0$ list.  The distance from a point to the $C_0$ list is defined to be the minimum distance between the point and elements of $C_0$. Thirdly, if there are any edge points left, we add the closest to the $C_0$ list and repeat the second step. The distance to the third step is recursively performed until no edge points are left.

The $R_0$ list for each $C_0$ is 5, 10, and 15 pc, and the corresponding angular size is calculated by assuming that the molecular cloud distance is $D_{\rm cut}/2$. For a molecular cloud at 500 pc, 15 pc is equivalent to 1\fdg7, too far from molecular clouds on the sky. Consequently, we restrict all off-cloud stars within  0\fdg5 of molecular cloud edge points.

Lengths of $\rm CO_{cut}$, $D_{\rm cut}$, and $R_0$ lists are 4, 7, and 3, respectively, while the list of $C_0$ depends on the sizes of molecular clouds. Since the $C_0$ list contains at least 2 points, the minimum number of parameter combinations is 168 (4$\times$7$\times$3$\times$2). The $C_0$ length grows fast for large molecular clouds, and to keep the total number of parameter combinations low, we require lengths of $\rm CO_{cut}$, $D_{\rm cut}$, and $R_0$ as short as possible. For each parameter combination, we select on- and off-cloud stars accordingly and estimate a distance with the method described in the next section.

\subsection{Fast Distance Estimate}
\begin{figure}[ht!]
\plotone{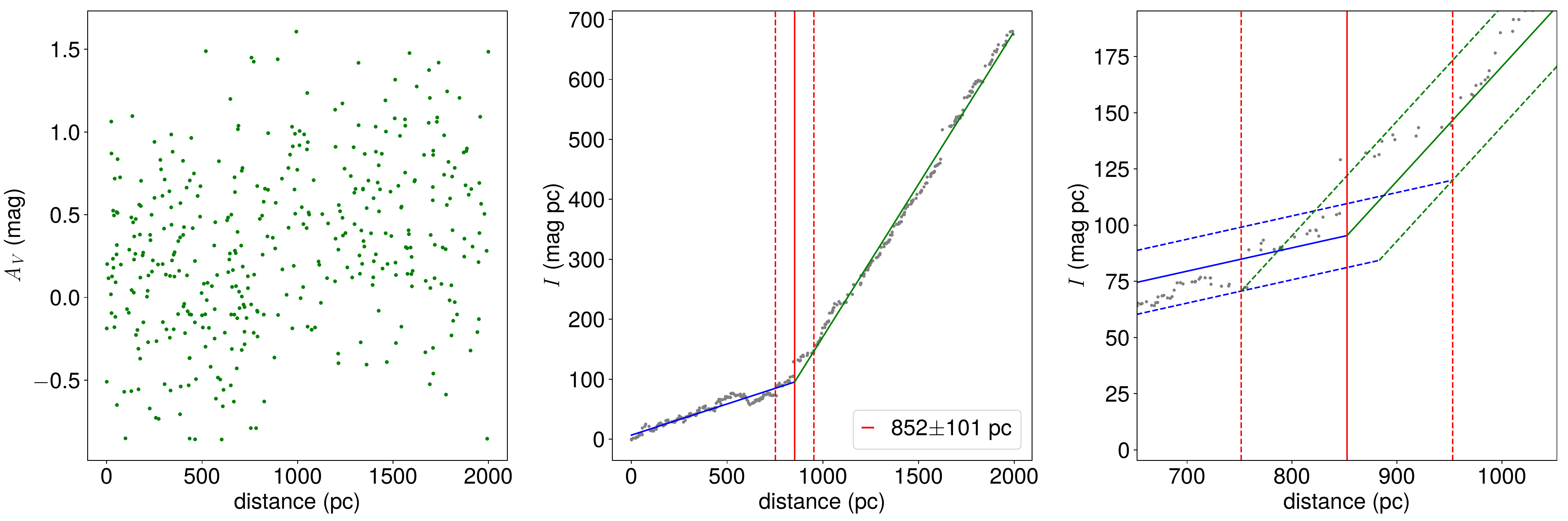} 
\caption{A demonstration of distance detection with the integrated extinction, $I$. This molecular cloud is simulated at 800 pc with foreground extinction of 0.1 mag and background extinction of 0.5 mag, both of which have a standard deviation of 0.5 mag. The left panel shows the variation of \av\ with respect to the distance. The middle panel plots values of $I$ derived with Equation \ref{eq:intFit}. The blue line represents the foreground part of $I$, while the green line represents the background part. The solid and dashed red lines mark the distance and error derived with $I$. The right panel zooms in the cross point in the middle panel, and presents the 1$\sigma$ range (vertically) with dashed blue and green lines, where $\sigma$ is the rms residual of the $I$ fitting. The distance error is defined to be the overlapping distance range of foreground and background $I$ strips with dashed blue and green lines as boundaries.  \label{fig:intSchem}}   
\end{figure}

In practice, the BEEP-II method hinges on fast and accurate distance estimates, because it examines a large number of parameter combinations. MCMC sampling, however, is too time-consuming to satisfy this requirement, so we propose to use piecewise linear regression to solve this problem.

Ideally, the foreground and background extinction of on-cloud stars show Gaussian distributions after removing the background extinction with off-cloud stars. The background extinction is expressed in a monotonically increasing function with respect to the distance to the Sun, and is subtracted from the on-cloud stellar extinction \citep{2019ApJ...885...19Y}. However, a direct piecewise linear regression (with slope zero) between the extinction and the distance of on-cloud stars usually yields incorrect results, due to the large dispersion of extinction.

An alternative approach is to integrate the extinction against the distance, making the foreground and background extinction-distance relationships more distinguishable. Specifically, given $N$ on-cloud stars, with distances, $D_i$ (in increasing order), and extinction, $A_i$ (with standard deviation $\Delta A_i$), where $1\leq i\leq N$, the integration interval for each $A_i$ is defined as
\begin{equation}
 \mathrm{d}D_i=\left\{
                \begin{array}{ll}
                   \frac{1}{2}\left(D_2-D_1\right), \  i = 1,\\
                   \frac{1}{2}\left(D_{i+1}-D_{i}\right)  +  \frac{1}{2}\left(D_{i }-D_{i-1}\right), \ 2 \leq i \leq N-1,\\
                    \frac{1}{2}\left(D_N-D_{N-1}\right), \  i = N,
                     \end{array}
              \right.
\label{eq:ddi}
\end{equation}
and the integrated extinction, $I_i$, is defined as 
\begin{equation}
 I_i =  \sum_{j=1}^{i}   A_j \mathrm{d}D_j,
\label{eq:intExt}
\end{equation}
where $1\leq i\leq N $ and $A_j$ is the stellar extinction with background extinction removed using off-cloud stars. $I_i$ is still a piecewise linear function with two steps, and slopes of the first and second segments correspond to the mean of foreground and background extinction. The standard deviation of $I_i$ is 
\begin{equation}
 \Delta I_i = \sqrt{ \sum_{j=1}^{i}  \left(\Delta  A_j  \mathrm{d}D_j \right)^2 }.
\label{eq:intExtError}
\end{equation}

The relative error of $\Delta I_i$  is usually larger than 50\%, much larger than the distance error of $D_i$, so only the error of $I_i$ is considered in piecewise linear regression. However, both the distance and extinction errors are taken into account in the step of MCMC sampling.

The effect of Equation \ref{eq:intExt} is demonstrated in Figure \ref{fig:intSchem}, using a molecular cloud at 800 pc. The foreground and background extinctions are 0.1 and 0.5 mag, respectively, both having standard deviations of 0.5 mag. The left-hand panel displays the distribution of $A_i$, while the middle panel plots values of $I_i$.

We use a two-step piecewise linear function to fit four parameters: $D$, $Y$, $\mu_1$, and $\mu_2$. $D$ and $Y$ represent the distance and the integrated extinction of the breakpoint, while $\mu_1$ and $\mu_2$ are the slopes of the foreground and background $I$, respectively. Based on the definition of $I$, $\mu_1$ and $\mu_2$ also represent average values of the foreground and background extinction, respectively. The specific form of $I_i$ is 
\begin{equation}
 I_i=\left\{
                \begin{array}{ll}
                  Y + \mu_1*(D_i-D), \ \mathrm{for} \ D_i\leq D,\\
                   Y + \mu_2*(D_i-D), \ \mathrm{for}\ D_i>D,
                     \end{array}
              \right.
\label{eq:intFit}
\end{equation}
where $D_i$ is the corresponding distance of $I_i$. We fit Equation \ref{eq:intFit}  with \texttt{curve\_fit} in the \texttt{Python} package of \texttt{scipy}.  As shown in Figure \ref{fig:intSchem}, the integrated extinction, $I$, reveals the distance accurately.

However, the error of $D$ given by \texttt{curve\_fit} is too small to represent true distance errors. Alternatively, we define a distance error based on the residual rms of the $I$ fitting. As shown in the right-hand panel of Figure \ref{fig:intSchem}, the distance error is defined as the overlapping distance range of foreground and background $I$ strips, which have vertical widths of 2$\sigma$, where $\sigma$ is the residual rms, different for foreground and background stars.


In this case of simulation, the BEEP-II method derives the molecular distance accurately, even for foreground and background star samples with large extinction dispersion. In the next section, we propose extra criteria to reject nonoptimal or incorrect distance results.

 \subsection{Rejecting False Distance Detections}
 \label{sec:reject}
 
  In this section, we describe the criteria used to reject nonoptimal parameters or incorrect distance estimates. The fast distance estimation method described in the previous section always provides solutions, but some may be false or nonoptimal. For instance, if a molecular cloud is at 900 pc, all distances produced with $\rm CO_{cut}$ 300 or 500 pc are incorrect. $\rm CO_{cut}$ 1000 pc may give correct distances, but $\rm CO_{cut}$ 1500 pc is obviously a better choice, because more background stars are used to describe the background extinction distribution.

The following criteria place constraints on $D$, $\mu_1$, $\mu_2$, and the extinction of off-cloud stars:
 \begin{enumerate} 
 \item The relative $D$ error defined in the right panel of Figure \ref{fig:intSchem} is at most 20\%. 
 \item If $D\leq \left(D_{\rm cut} - 1500\right) \ \rm pc$, this parameter combination is rejected, because there is a better one, which is $\left(D_{\rm cut} -500\right)$ pc. To keep on- and off-cloud stars close to each other, $D_{\rm cut}$ should not be too large.  
 
  \item   If $D> \left(D_{\rm cut}-100\right) \ \rm pc$ and $D_{\rm cut}<3000 \ \rm pc$, this parameter combination is rejected, because there is a better one, which is $\left(D_{\rm cut} +500\right)$ pc. $D_{\rm cut}$ should be large enough to include sufficient background stars. 

   \item   If $D< 200 \ \rm pc$ and $D_{\rm cut}> 300 \ \rm pc$, this parameter combination is rejected, because there is a better one, which is 300 pc.

   \item  If $\mu_1$ is larger than $\mu_2$, this parameter combination is rejected. This criterion means that the foreground extinction should be smaller than the  background extinction. 
   
   \item The number of background off-cloud stars is required to be at least half the number of background on-cloud stars. This criterion is used to reject regions where off-cloud stars are insufficient to trace the extinction variation near molecular clouds.
   
   \item The Gaussian distributions of the extinction of background and foreground on-cloud stars are separated from each other, as well as those of background  on-cloud stars and background off-cloud stars. The separation of two Gaussian distributions, $\mathcal{N}(A_1,\,\sigma_1 )$ and $\mathcal{N}(A_2,\,\sigma_2 )$, requires $|A_1-A_2|>\sigma_2$ or $|A_1-A_2|>\sigma_1$, where $A_1$ and $A_2$ are the means of two Gaussian distributions and $\sigma_1$ and $\sigma_2$ are their standard deviations.
\end{enumerate}

Those seven criteria effectively reject false detections. We cluster remaining distance results and select optimal parameters in the next section.

\subsection{Clustering}
\label{sec:clustering}

 The clustering operation is used to determine distance components of molecular clouds. For small molecular clouds, this refers to distance components along the line of sight, while for large molecular clouds, this also includes components that are far from each other across the sky. In addition, the clustering significantly reduces the number of parameter combinations fed to MCMC sampling.

The clustering is done in the 3D physical space, calculated with $C_0$ and distances of circular regions. The X axis corresponds to $\left(l, b\right) = \left(90\deg, 0\deg\right)$, while the Y axis corresponds to $\left(l, b\right) = \left(180\deg, 0\deg\right)$. The Z axis is perpendicular to the X-Y plane, i.e., $b=90\deg$. The algorithm we used is a hybrid of DBSCAN and k-means. DBSCAN is efficient in identifying outliers, while k-means is useful to split samples into a desired number of parts. 

DBSCAN has two parameters, eps and MinPts. Eps specifies the maximum distance between two neighbors in the above-mentioned 3D physical space. MinPts is the minimum number of neighbours for a point to form a dense region, and dense regions together with their neighbors form clusters. Because DBSCAN is only used to identify evident outliers, we adopt loose parameters: 200 pc for eps and 3 for MinPts. Distances not included in DBSCAN clusters are considered outliers. 

After removing obvious outliers with DBSCAN, we cluster the remaining molecular cloud distances with k-means, weighted by the reciprocal of the distance variance. K-means splits data points by minimizing the variance with respect to a certain number of centroids. The problem of k-means is that we have to specify the total number of clusters, which is unknown.

However, for molecular cloud distances, we can set criteria to prevent them from being split into too many small pieces, and we require that the coordinate differences between cluster centroids in all three axes are larger than 20\% and 10 pc. 20\% follows the parallax error threshold of Gaia DR2 stars, and 10 pc is the typical size of molecular clouds. Specifically, we adopted a binary-splitting process to generate k-means clusters. Starting from one single cluster that includes all distances, existing clusters are recursively divided into two. If no clusters can be split without violating the above-mentioned criteria, k-means stops.

\subsection{The Optimal Parameter Combination}
\label{sec:optimal}

For each cluster produced in Section \ref{sec:clustering}, we select one optimal parameter combination (i.e. a combination of $D_{\rm cut}$, $\rm CO_{cut}$, $C_0$, and $R_0$) corresponding to a distance derived with a pair of  on- and off-cloud star samples. Because the distance error is required to be small, we reject half of the parameter combinations that yield the largest distance errors. The optimal parameter combination is selected from the remaining half. 

 
In general, the optimal parameter combination of a cluster is selected by maximizing the number of foreground ($\leq D$) or background ($> D$) on-cloud stars, depending on whichever has fewer stars. Specifically, if the mean (over all cases in a cluster) number of foreground on-cloud stars is less than half the mean number of background on-cloud stars, the optimal parameter combination is the one that has the largest number of foreground on-cloud stars. Likewise, if the mean number of foreground on-cloud stars is larger than half the number of background on-cloud stars, the optimal parameter combination is the one that has the largest number of background on-cloud stars. However, when numbers of foreground and background on-cloud stars are balanced, i.e., the difference is less than half, we choose the parameter that maximizes the total number of on-cloud stars. 

If the foreground or background on-cloud stars are abundant, a smaller $D_{\rm cut}$, a smaller $R_0$, or a higher $\rm CO_{\rm cut}$  is tried. The abundance of on-cloud stars means that at least one parameter combination yields a total number of on-cloud stars exceeding 100, with the number of foreground and background on-cloud stars each exceeding 50. In this scenario, we try stricter parameter combinations which include fewer on-cloud stars, but with less contamination of background stars (a smaller $D_{\rm cut}$),  more accurate background extinction elimination (a smaller $R_0$), or clearer jumps of extinction (a higher $\rm CO_{\rm cut}$).
 
If on-cloud stars are abundant, we use a trial-and-error approach to obtain better parameter combinations.  Specifically, we keep removing the second largest $D_{\rm cut}$ until on-cloud stars will be no longer abundant after removing parameter combinations with the maximum $D_{\rm cut}$. Similar procedures are applied to $R_0$ and $\rm CO_{\rm cut}$, and the trial sequence of parameters is $D_{\rm cut}$,  $R_0$, and $\rm CO_{\rm cut}$. Because $D_{\text{cut}}$ and $R_0$ have a less significant effect on the on-cloud stars sample, they are adjusted before  $\rm CO_{\rm cut}$. This trial-and-error approach only operates on cases with abundant on-cloud stars, and   at least one abundant case remains afterward.


We impose an extra criterion on $D_{\rm cut}$ to prevent the process of rejecting abundant cases from going too far. In all rejected cases, $D_{\rm cut}$  is required to be at least 1.5$\times$$D_{\rm avg}$ pc, where $D_{\rm avg}$ is the average distance (weighted by the inverse distance variance) of the cluster.

\begin{deluxetable}{ccccccccc}
\tablecaption{Statistics of Molecular Cloud Distances in Three Regions\tablenotemark{a}.  \label{Tab:disThisWork}}
\tablehead{
\colhead{Quadrant} & \colhead{$l$} & \colhead{$b$}  &  \colhead{$V_{\rm LSR}$}   & \colhead{Algorithm} & \colhead{Total Number\tablenotemark{b} } & 
\colhead{Measured Clouds} &  \colhead{Cloud Distances}  \\
\colhead{   } & \colhead{(deg)} & \colhead{(deg)} & (km/s) &  &    & 
}
\startdata
First  &  [25.8, 49.7] &  [-5, 5]& [-6, -30]& DBSCAN & 359   &  27  &  27    \\ 
Second  &  [104.75, 150.25] &  [-5.25, 5.25]& [-95, 25]& DBSCAN  & 1618   &   155   &   159\tablenotemark{c} \\ 
Third  &  [209.75, 219.75] &  [-5, 5]& [0, 70]& DBSCAN & 237   &  52  & 52   \\ 
\enddata
\tablenotetext{a}{See Table \ref{Tab:disMWISP} for a comparison with previous BEEP results.}
\tablenotetext{b}{This only takes into account molecular clouds that are complete in the PPV space with angular areas larger than 0.015 deg$^2$.}
\tablenotetext{c}{One clouds has multiple distances.}
\end{deluxetable}


\begin{figure}[ht!]
 \gridline{
 \fig{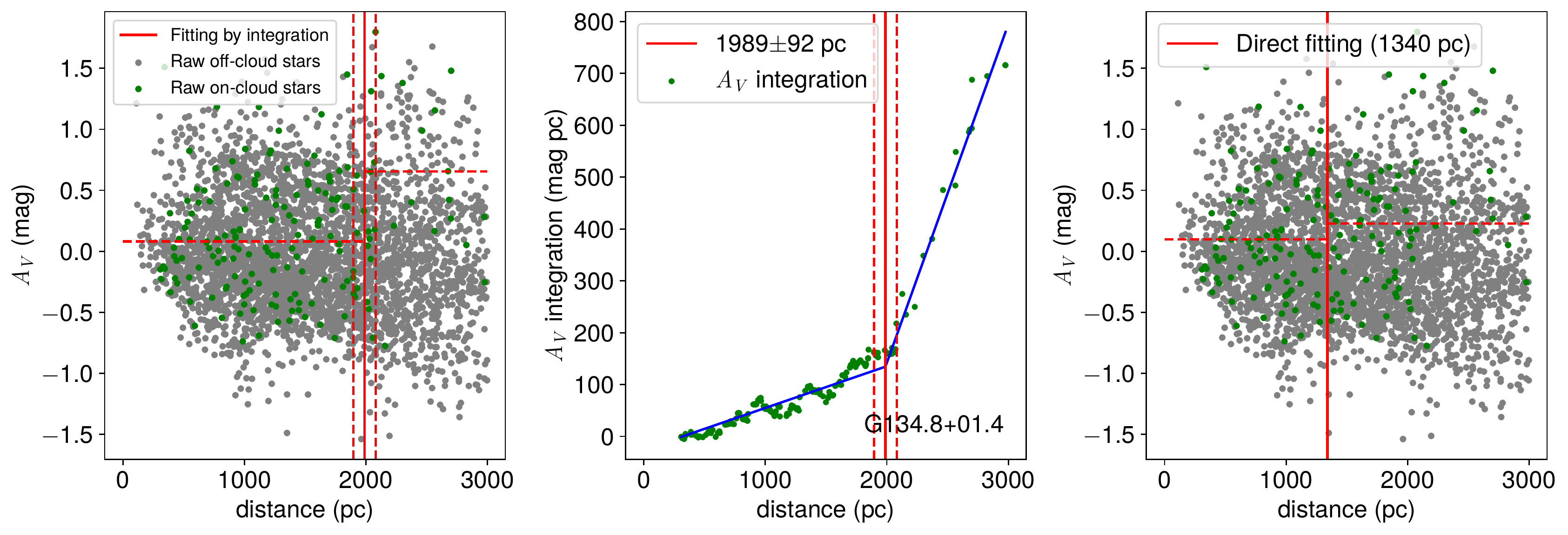}{0.9\textwidth}{ (a) }  
 } 
  \gridline{
 \fig{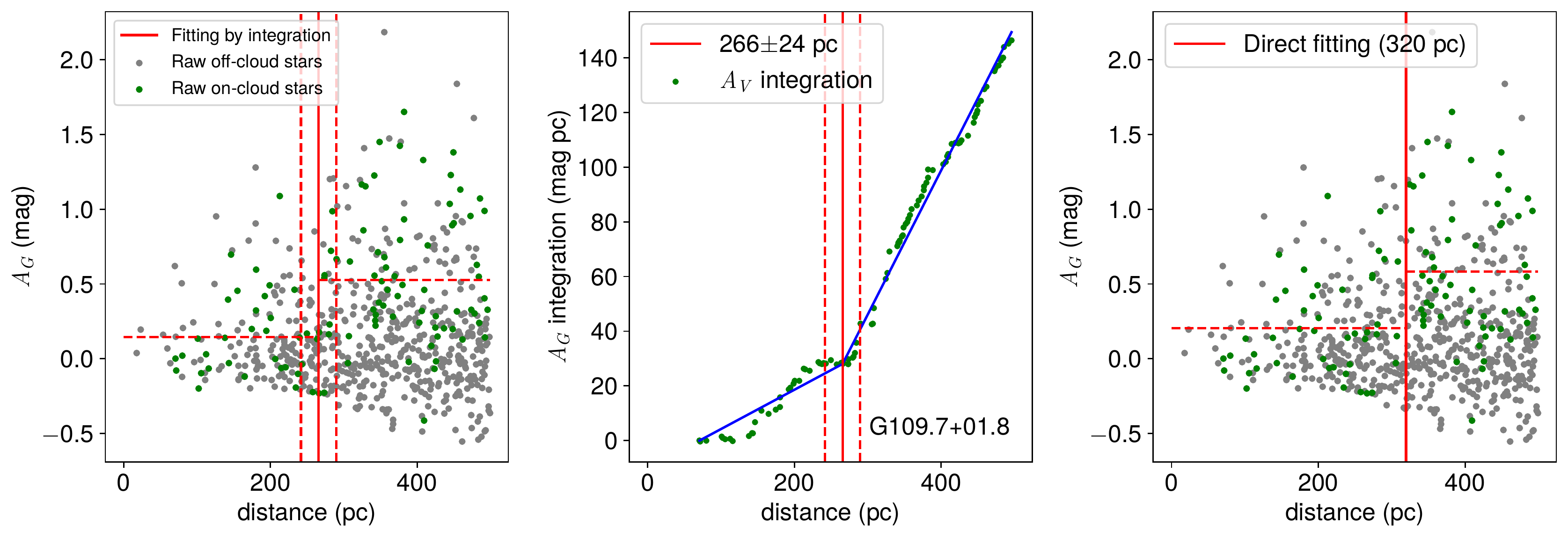}{0.9\textwidth}{(b)}  
 } 
\caption{Fast distance estimate with the BEEP-II method in regions of (a) \wfour\ and (b) \largestMC. Green and gray dots represent raw on- and off-cloud stars with the  extinction background removed, respectively. Blue lines indicate piecewise linear regression of \av\ integration according to Equation \ref{eq:intFit}. Distances derived from  \av\ integration are displayed in the left and middle panel, and as a comparison, a direct fitting of foreground and background extinction is shown in the right panel. \label{fig:fastestimate} }
\end{figure}

 
 %

The optimal parameter combination is used to derive the distance of each cluster, and the extinction jump is required to be clear. If the distance to a molecular cloud is measured using both $A_G$ and $A_V$, the one with a smaller distance error or with a clearer extinction jump was selected. The selection of good MCMC sampling results and the choice of $A_G$ or $A_V$ distances is judged by eye.

\begin{figure}[ht!]
 \gridline{
 \fig{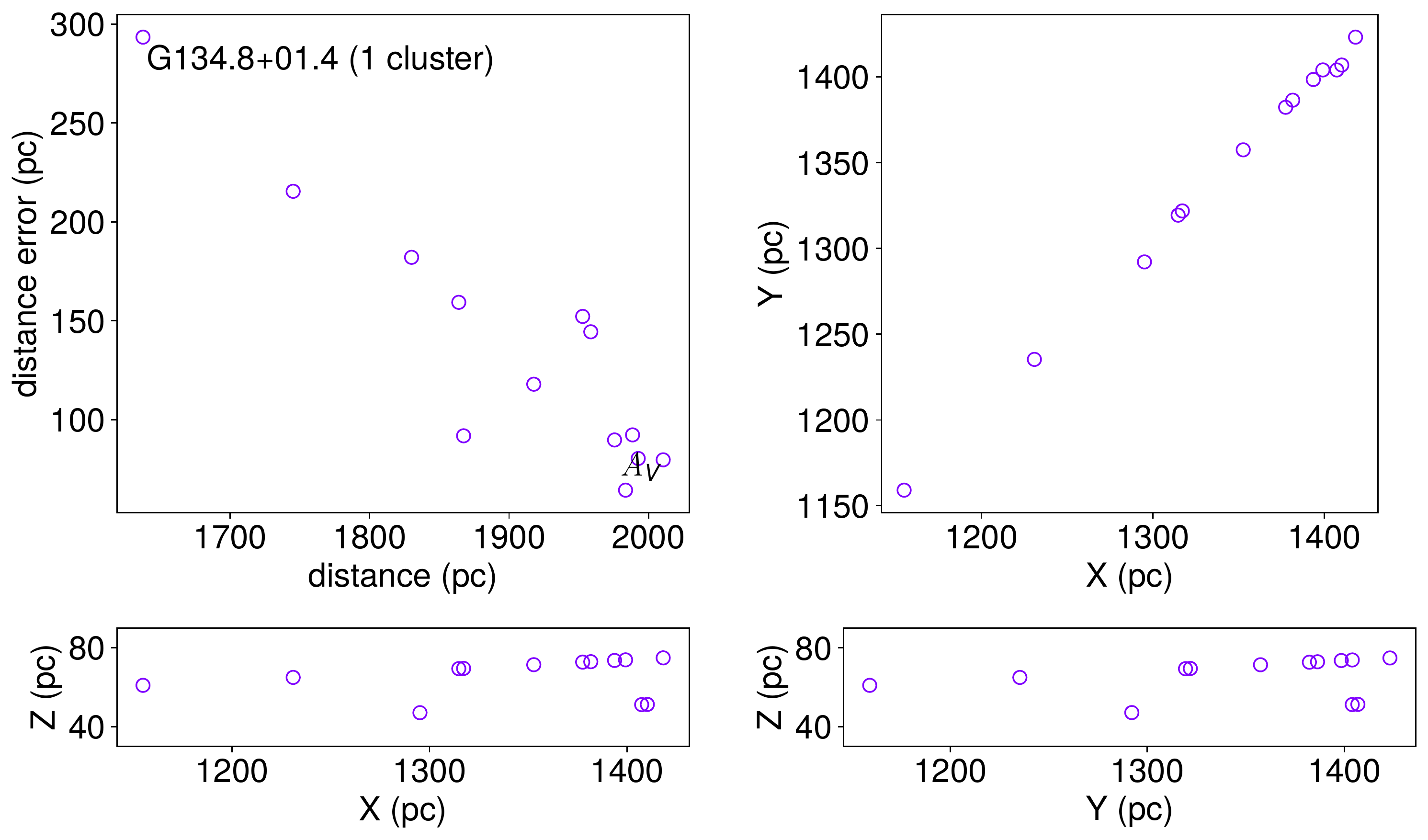}{0.49\textwidth}{(a) }    \fig{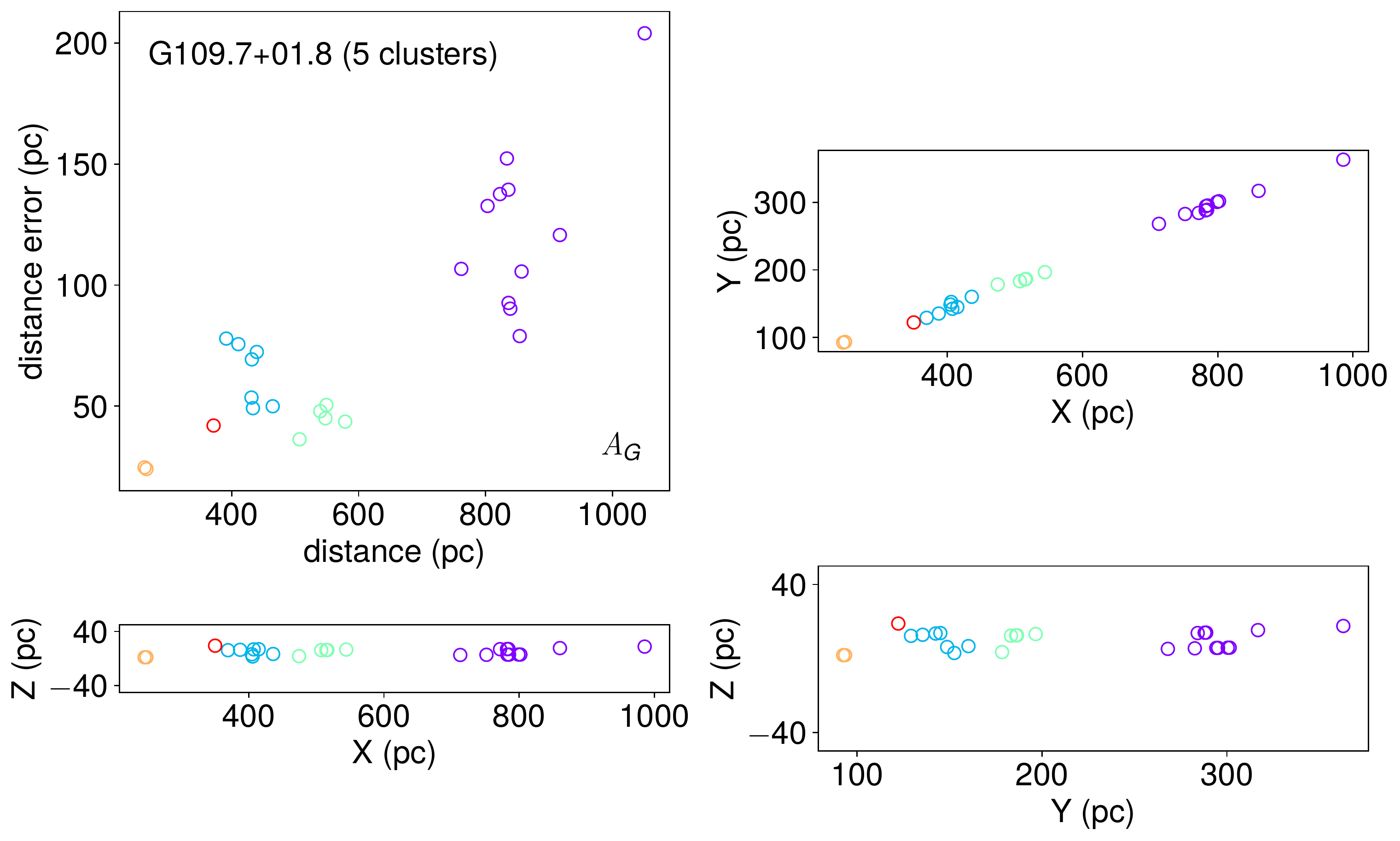}{0.49\textwidth}{(b)}  
 } 

\caption{Clustering of (a) \wfour\ with \av\ and (b) \largestMC\ with \ag. For each molecular cloud, the above left panel displays the distance and error, and the other three panels show the 2D projection of clusters in three planes. The coordinate origin is placed at the location of the Sun, and colors represent distinct clusters. The X axis corresponds to $l=90^\circ$, while the Y axis corresponds to $l=180^\circ$. The Z axis is perpendicular to the X-Y plane.  \label{fig:cluster} } 
\end{figure}

\begin{figure}[ht!]
 \gridline{
 \fig{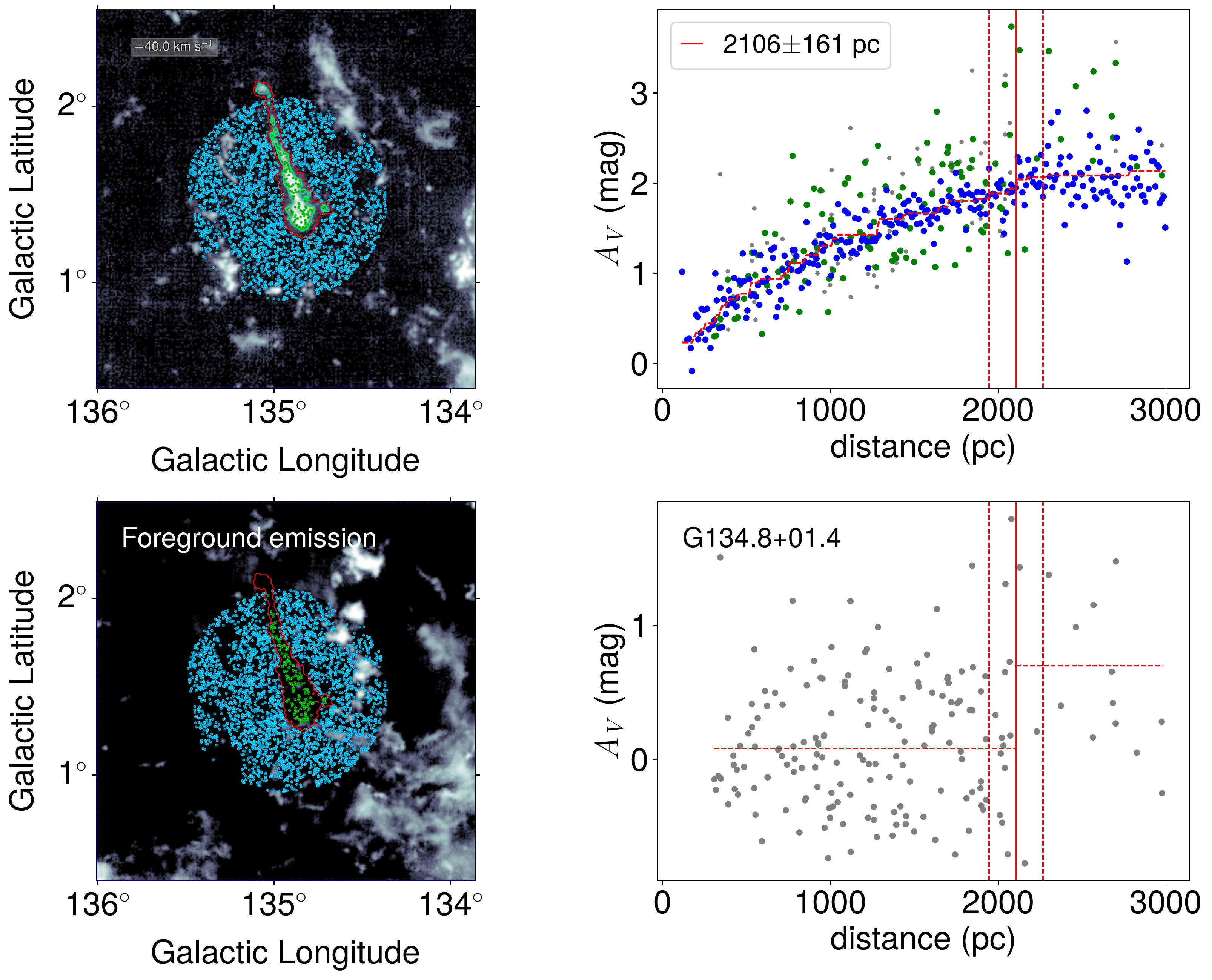}{0.49\textwidth}{(a)}   \fig{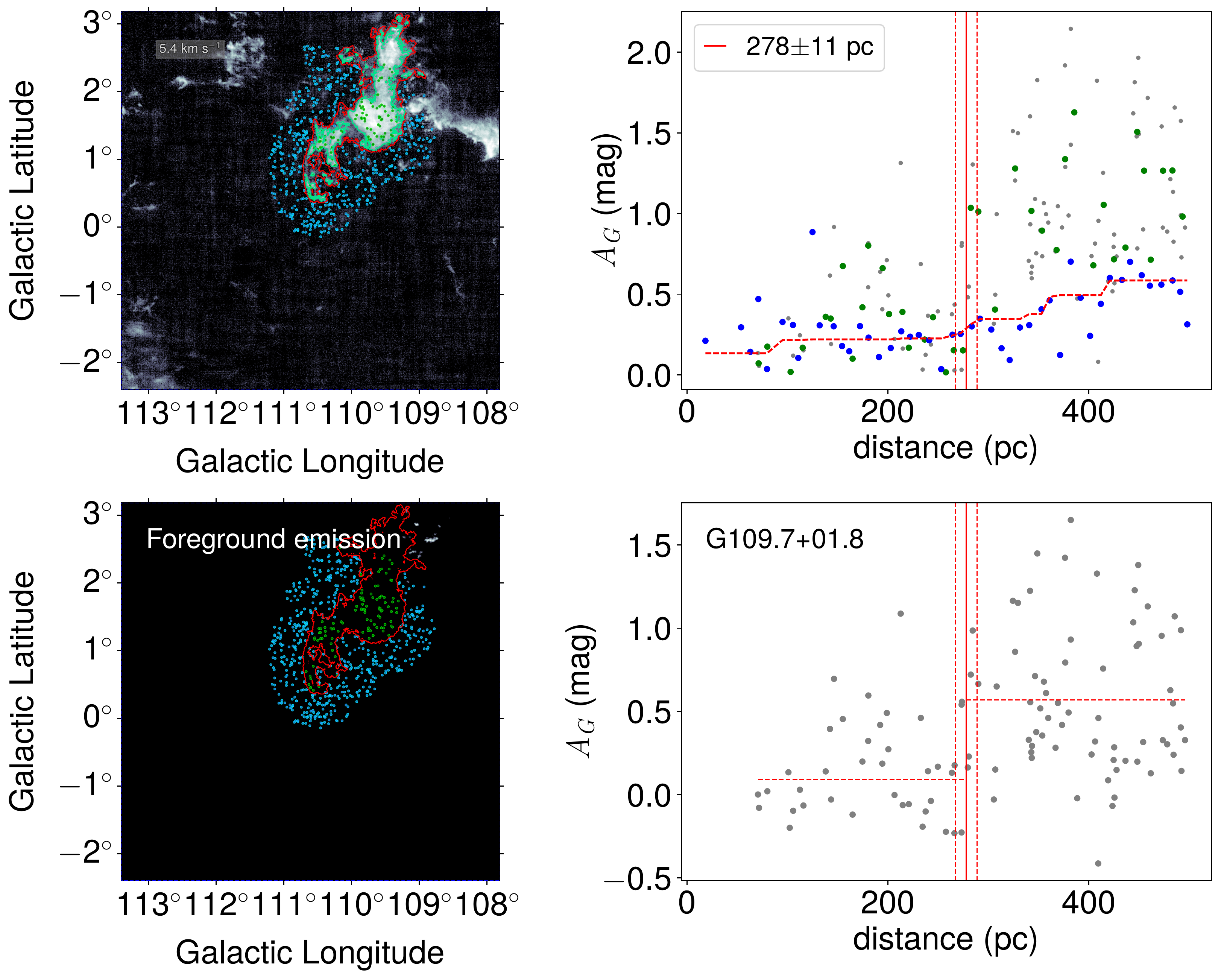}{0.49\textwidth}{(b)}  
 } 

\caption{Distances of (a) \wfour\ and (b) \largestMC\ derived with MCMC sampling using parameters provided by the BEEP-II method. The left-hand side shows the integrated images (top) of molecular clouds and their foreground emission (bottom), as well as positions of on- (green) and off-cloud (cyan) stars. In the top right panel, green dots represent binned on-cloud stars, while blue dots represent binned off-cloud stars. Gray dots in the top right panel represent raw on-clouds, while in the bottom right panels, they represent background-removed on-cloud stars. \label{fig:mcmc}}   
\end{figure}

\subsection{Removing Foreground Detections}
\label{sec:foreground}

The foreground emission is defined with radial velocity. Denoting the weighted average (by intensity) and the weighted standard derivation of the radial velocity with $V$ and $\Delta V$, respectively, and letting  $V_0$  be the local velocity (close to 0 \kms\ but changing with the Galactic longitude), the integration velocity range of foreground emission is from  $V_0$ to  $\left(V+3\Delta V\right)$ or $\left(V-3\Delta V\right)$, depending on which one is closer to $V_0$. It is possible that the foreground emission defined in this way is actually background emission, but this does not adversely affect the distance measurement because removing stars towards those molecular clouds only makes the background extinction more closely fit a Gaussian distribution.

In both the BEEP and the BEEP-II methods, we remove stars that are severely ($>$ 3 K \kms) affected by foreground emission, but distances to foreground dust can still be measured under certain favorable conditions, in the presence of high dust extinction near large CO clouds.  Despite low CO emission, those regions may cause significant extinction, and if a distant molecular cloud is occasionally located behind those regions, distances to its foreground emission are often measurable.

To remove foreground detections, we examine the extinction variation toward foreground emission with the BEEP method. For foreground molecular clouds, we distinguish on- and off-cloud stars with only CO integrated intensity  $>$ 4 K \kms\ for on-cloud stars and $<$ 1 K \kms\ for off-cloud stars, keeping  other parameters the same for targeting molecular clouds.


\begin{deluxetable}{ccccccccccccc}
\setlength{\tabcolsep}{3.5pt}
\tabletypesize{\scriptsize}
\tablecaption{Distances to Molecular Clouds.\label{Tab:discat}}
\tablehead{
   Name &  $l$  &   $b$   &\colhead{$V_{\rm LSR}$} &  Area &  $D_{\rm Gaia}$\tablenotemark{a} &  Region Center\tablenotemark{b} &  Radius\tablenotemark{b}   & CO$_{\rm cut}$\tablenotemark{c}   &  $D_{\rm cut}$   & N\tablenotemark{d} &  Mass\tablenotemark{e}        &$A_V$/$A_G$   \\
   &  (\deg)  &  (\deg)  & (km/s)  &  deg$^2$ &   (pc)   & (\deg,\deg)  & (\deg)   & (K km/s) & (pc) &  & ($10^3$ \msun)           
}
\colnumbers
\startdata 
G026.9-03.5 &  26.913 & -3.571 &   16.3 &   0.33 & $ 445\pm 17$ & ( 26.533,-3.992) & 10.00 & 3.0 & 1000 &   239 & $   0.45\pm  0.06$ & $A_V$ \\
G027.8-02.1 &  27.893 & -2.170 &   18.0 &   1.92 & $ 484\pm  8$ & ( 27.508,-2.333) & 10.00 & 5.0 & 1000 &   407 & $   9.34\pm  0.98$ & $A_G$ \\
G028.6-02.9 &  28.635 & -2.946 &   18.5 &   0.04 & $ 826\pm 94$ & ( 28.633,-2.950) & 10.00 & 2.0 & 2000 &    69 & $   0.08\pm  0.02$ & $A_V$ \\
G032.9-01.8 &  32.954 & -1.842 &   22.4 &   0.02 & $2392\pm192$ & ( 32.950,-1.842) &  5.00 & 2.0 & 3000 &    61 & $   0.36\pm  0.07$ & $A_V$ \\
G034.2-01.7 &  34.254 & -1.707 &    6.6 &   0.14 & $ 532\pm 61$ & ( 33.942,-1.608) & 10.00 & 2.0 & 2000 &   164 & $   0.25\pm  0.06$ & $A_G$ \\
G037.0+04.6 &  37.053 &  4.660 &   22.8 &   0.04 & $1016\pm 85$ & ( 37.067, 4.667) &  5.00 & 3.0 & 1500 &    30 & $   0.19\pm  0.04$ & $A_V$ \\
G038.1-01.2 &  38.159 & -1.227 &    7.7 &   0.20 & $ 698\pm 64$ & ( 38.158,-1.225) &  5.00 & 3.0 & 1000 &    63 & $   0.37\pm  0.08$ & $A_V$ \\
G038.4-03.3 &  38.432 & -3.388 &    7.5 &   0.11 & $ 872\pm 66$ & ( 38.467,-3.417) & 10.00 & 2.0 & 1500 &    94 & $   0.26\pm  0.05$ & $A_V$ \\
G039.4-02.7 &  39.454 & -2.741 &   15.0 &   0.21 & $ 493\pm 20$ & ( 39.433,-2.750) & 10.00 & 3.0 & 1500 &   235 & $   0.30\pm  0.04$ & $A_V$ \\
G041.5+02.3 &  41.581 &  2.328 &   17.6 &   1.11 & $ 909\pm 13$ & ( 41.550, 2.625) & 10.00 & 5.0 & 1500 &   518 & $   10.4\pm   1.1$ & $A_G$ \\
G041.5-01.8 &  41.583 & -1.820 &   16.8 &   0.14 & $ 708\pm 65$ & ( 41.458,-1.550) & 15.00 & 2.0 & 2000 &    94 & $   0.38\pm  0.08$ & $A_G$ \\
G041.9+02.2 &  41.981 &  2.300 &    5.4 &   1.75 & $ 314\pm 15$ & ( 41.542, 2.633) & 10.00 & 3.0 & 1000 &   390 & $   1.09\pm  0.15$ & $A_G$ \\
G042.1-02.1 &  42.160 & -2.105 &   23.4 &   0.06 & $1066\pm 55$ & ( 42.167,-2.100) & 10.00 & 2.0 & 1500 &    53 & $   0.48\pm  0.07$ & $A_G$ \\
G042.9+04.4 &  42.975 &  4.423 &   22.7 &   0.02 & $1441\pm 68$ & ( 42.967, 4.425) & 10.00 & 2.0 & 2500 &    89 & $   0.18\pm  0.02$ & $A_V$ \\
G043.3+03.1 &  43.386 &  3.154 &   10.2 &   0.22 & $ 723\pm 16$ & ( 43.450, 3.100) &  5.00 & 4.0 & 1000 &   153 & $   0.98\pm  0.11$ & $A_V$ \\
G043.4-04.2 &  43.427 & -4.273 &   26.1 &   0.03 & $1021\pm 52$ & ( 43.483,-4.142) & 10.00 & 2.0 & 2500 &   109 & $   0.13\pm  0.02$ & $A_V$ \\
G044.5+02.7 &  44.553 &  2.707 &   14.6 &   0.17 & $ 621\pm 16$ & ( 44.500, 2.808) &  5.00 & 2.0 & 1000 &   162 & $   0.37\pm  0.04$ & $A_V$ \\
G044.7-04.2 &  44.757 & -4.203 &   22.8 &   0.03 & $1154\pm166$ & ( 44.800,-4.175) & 15.00 & 2.0 & 2000 &    73 & $   0.20\pm  0.06$ & $A_V$ \\
G044.7+04.0 &  44.760 &  4.035 &   20.0 &   0.85 & $ 829\pm 13$ & ( 45.308, 4.092) &  5.00 & 4.0 & 1500 &   246 & $   4.82\pm  0.50$ & $A_V$ \\
G045.1-03.4 &  45.166 & -3.435 &   20.7 &   0.07 & $ 923\pm 72$ & ( 45.167,-3.442) & 15.00 & 2.0 & 2000 &   125 & $   0.18\pm  0.03$ & $A_V$ \\
G045.4-04.2 &  45.478 & -4.271 &   18.8 &   0.40 & $ 931\pm 25$ & ( 45.383,-4.217) &  5.00 & 3.0 & 1500 &   206 & $   2.56\pm  0.29$ & $A_G$ \\
G045.5-03.3 &  45.541 & -3.331 &   20.2 &   0.07 & $ 873\pm 41$ & ( 45.275,-3.483) & 10.00 & 2.0 & 2000 &   172 & $   0.24\pm  0.03$ & $A_V$ \\
G046.9-04.8 &  46.956 & -4.889 &   15.7 &   0.04 & $ 841\pm 45$ & ( 46.975,-4.942) & 10.00 & 2.0 & 1500 &    59 & $   0.10\pm  0.02$ & $A_V$ \\
G046.9+01.7 &  46.978 &  1.774 &    6.8 &   0.46 & $ 491\pm103$ & ( 47.042, 1.683) & 10.00 & 2.0 & 1000 &   157 & $   0.63\pm  0.27$ & $A_G$ \\
G047.3-03.1 &  47.353 & -3.175 &   18.7 &   0.04 & $ 912\pm 99$ & ( 47.375,-3.200) &  5.00 & 2.0 & 2000 &   151 & $   0.24\pm  0.06$ & $A_V$ \\
G047.3-03.0 &  47.355 & -3.035 &   24.4 &   0.03 & $1127\pm 88$ & ( 47.400,-3.058) & 10.00 & 2.0 & 2000 &    36 & $   0.18\pm  0.03$ & $A_G$ \\
G049.4-04.6 &  49.413 & -4.626 &   18.7 &   0.02 & $ 957\pm 96$ & ( 49.517,-4.642) & 15.00 & 2.0 & 2500 &    86 & $   0.10\pm  0.02$ & $A_V$ \\
... & ... & ...&   ... &  ... &  ... & ... &  ... & ... & ... &   ... &    ... & ... \\ 
\enddata
\tablenotetext{a}{The distance error is the standard deviation, and the 5\% systematic error is not included. Only molecular cloud distances in the first Galactic quadrant are displayed, and a machine readable table and all figures of all 238 distances of 234 molecular clouds are publicly accessible on the Harvard Dataverse (\href{https://doi.org/10.7910/DVN/0J76GM}{https://doi.org/10.7910/DVN/0J76GM}).} 
\tablenotetext{b}{Circular regions that contain on- and off-cloud stars.} 
\tablenotetext{c}{The lower threshold of CO emission for on-cloud stars.} 
\tablenotetext{d}{Total number of on-cloud stars.} 
\tablenotetext{e}{The mass estimate only takes into account CO-bright molecular gas, and only the distance error is considered, including the 5\% systematic error. Only molecular clouds with unique distances have masses calculated.} 
\end{deluxetable}


If the extinction jump of a molecular cloud and that of its foreground emission are close, we cannot judge which distance is measured, and so we exclude this kind of molecular clouds from our sample. In addition, if the angular size of a molecular cloud is overwhelmingly small compared with the foreground emission, its distance is unreliable, and this cloud is also excluded.

\section{Results}
\label{sec:result} 
 In this section, we display distances derived with the BEEP-II method in three regions listed in Table \ref{Tab:disMWISP}. As examples, we display results of \wfour\ and \largestMC. \wfour\ is located in the  W4 region \citep[e.g.,][]{1998ApJ...502..265H,2020ApJS..246....7S}, while \largestMC\ is a molecular cloud close to the Sun, both in the second Galactic quadrant.

\subsection{Fast Distance Estimate}

Figure \ref{fig:fastestimate} displays distances measured with piecewise linear regression with integrated extinction in the BEEP-II method. As a comparison, we show direct fitting results with piecewise linear regression (slopes are set to zero) using the extinction data. In both cases, the direct fitting approach does not gauge the right extinction jump, and in contrast, the BEEP-II method derives distances accurately. Figure \ref{fig:fastestimate} demonstrates that the BEEP-II method works well even when the numbers of foreground and background stars are unbalanced, provided that the extinction jump is evident.



\begin{figure}[ht!]
 \gridline{
 \fig{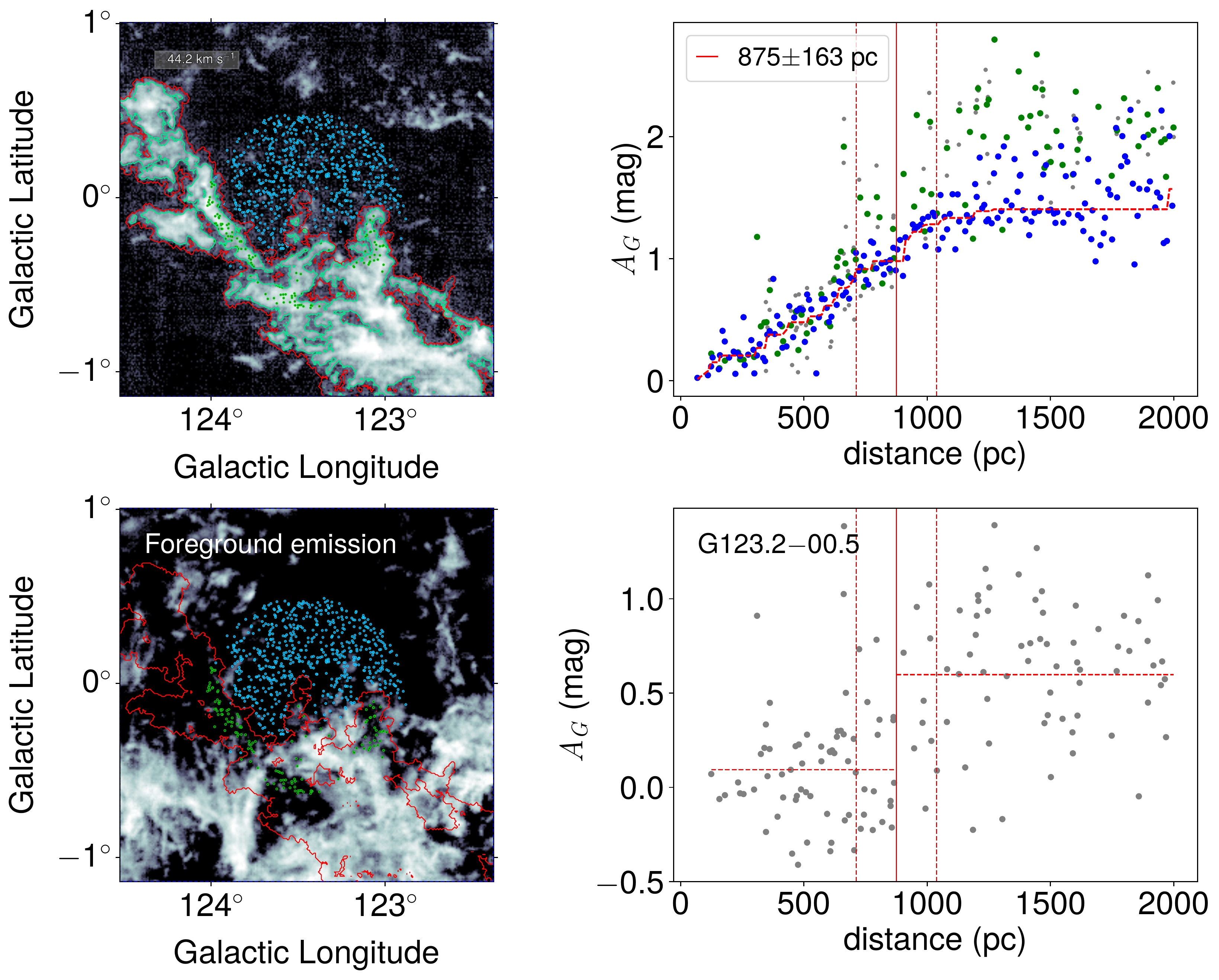}{0.49\textwidth}{(a)}   \fig{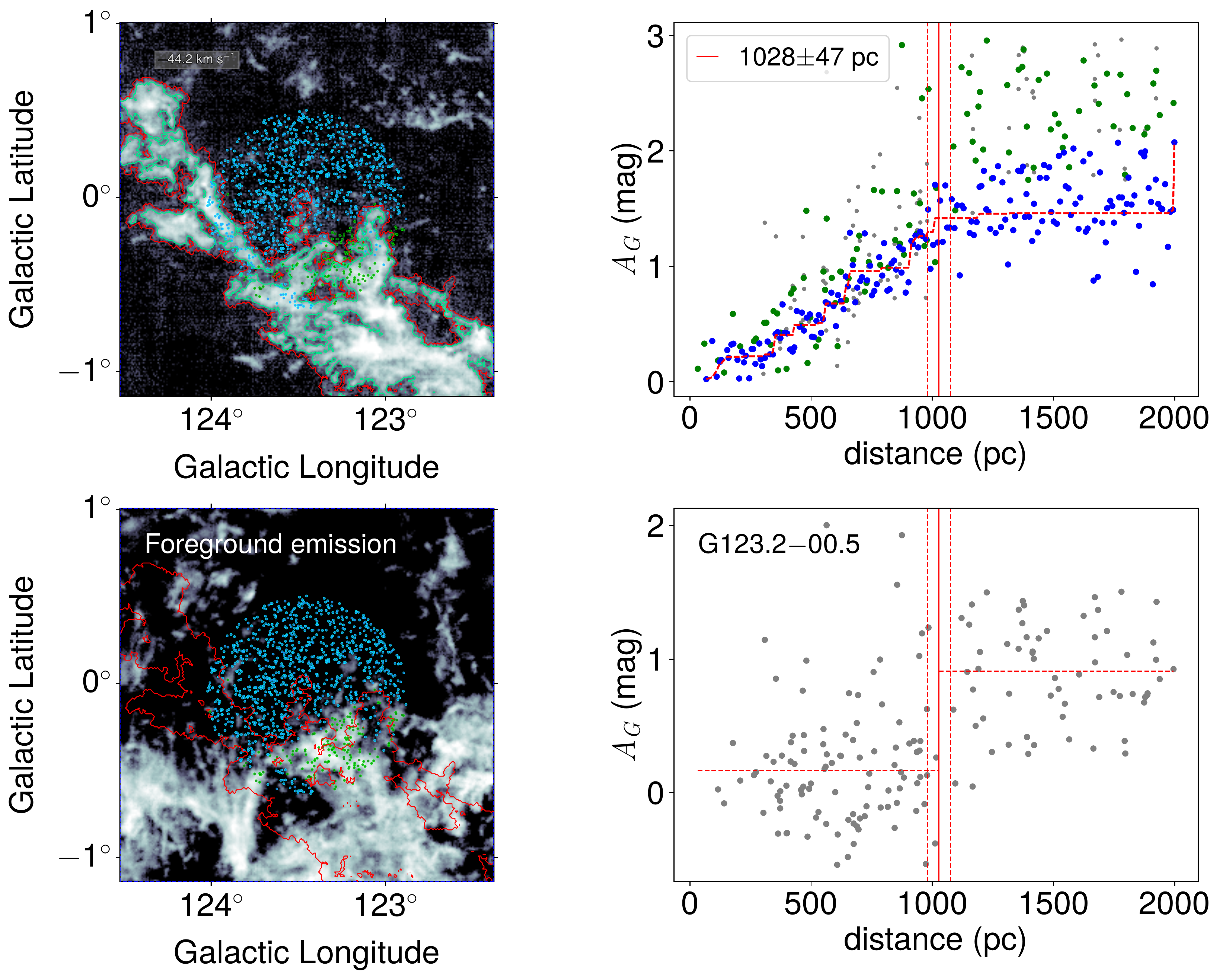}{0.49\textwidth}{(b)}  
 } 
 
\caption{Demonstration of the foreground ambiguity. Distances of (a) G123.2$-$00.5 and (b) its foreground are close, causing ambiguity. In panel (b), only the  brightness of CO is used to classify on- and off-cloud stars for foreground emission. See Figure \ref{fig:mcmc} for other details.   \label{fig:foreground} } 
\end{figure}

\subsection{Clustering and MCMC Sampling}

 Figure \ref{fig:cluster} shows the clustering results of \wfour\ and \largestMC. One cluster (\av) was found for \wfour, while five clusters (\ag) were found for \largestMC. The clustering procedure proposed in Section \ref{sec:clustering} works well in both sparse and crowded environments.

Each circle in Figure \ref{fig:cluster} represents a distance derived with piecewise linear regression, corresponding to a specific parameter combination.  For each cluster, we selected an optimal parameter combination with the method described in Section \ref{sec:optimal} and ran MCMC sampling accordingly.  
 
Figure \ref{fig:mcmc} displays MCMC sampling results for the clusters found for \wfour\ and \largestMC\ (one of its five clusters with \ag).  For each panel, the left-hand side displays the integrated (top) and foreground (bottom) images, and the right-hand side shows raw (top) and background-removed (bottom) extinction of on-cloud stars. For both molecular clouds, MCMC sampling derives distances accurately.

\subsection{The Distance Catalog}

We examine MCMC results of each cluster and only keep those with evident extinction jumps. If two circular regions largely overlap, only one is kept, usually the one with a smaller error. As described in Section \ref{sec:foreground}, if the foreground emission of molecular cloud shows an extinction jump at a close distance, this molecular cloud is excluded due to the foreground ambiguity.

An overview of the distance results is described in Table \ref{Tab:disThisWork}. In total, we derived distances to 234 molecular clouds, including 238 positions. The BEEP-II method allowed us to measure many more distances than the original BEEP method.

The distance catalog is displayed in Table \ref{Tab:discat}.  Table \ref{Tab:discat} only contains molecular clouds in the first Galactic quadrant, and the full catalog and all distance figures are publicly accessible on the Harvard Dataverse (\href{https://doi.org/10.7910/DVN/0J76GM}{https://doi.org/10.7910/DVN/0J76GM}).

\begin{figure}[ht!]
 
 \gridline{
 \fig{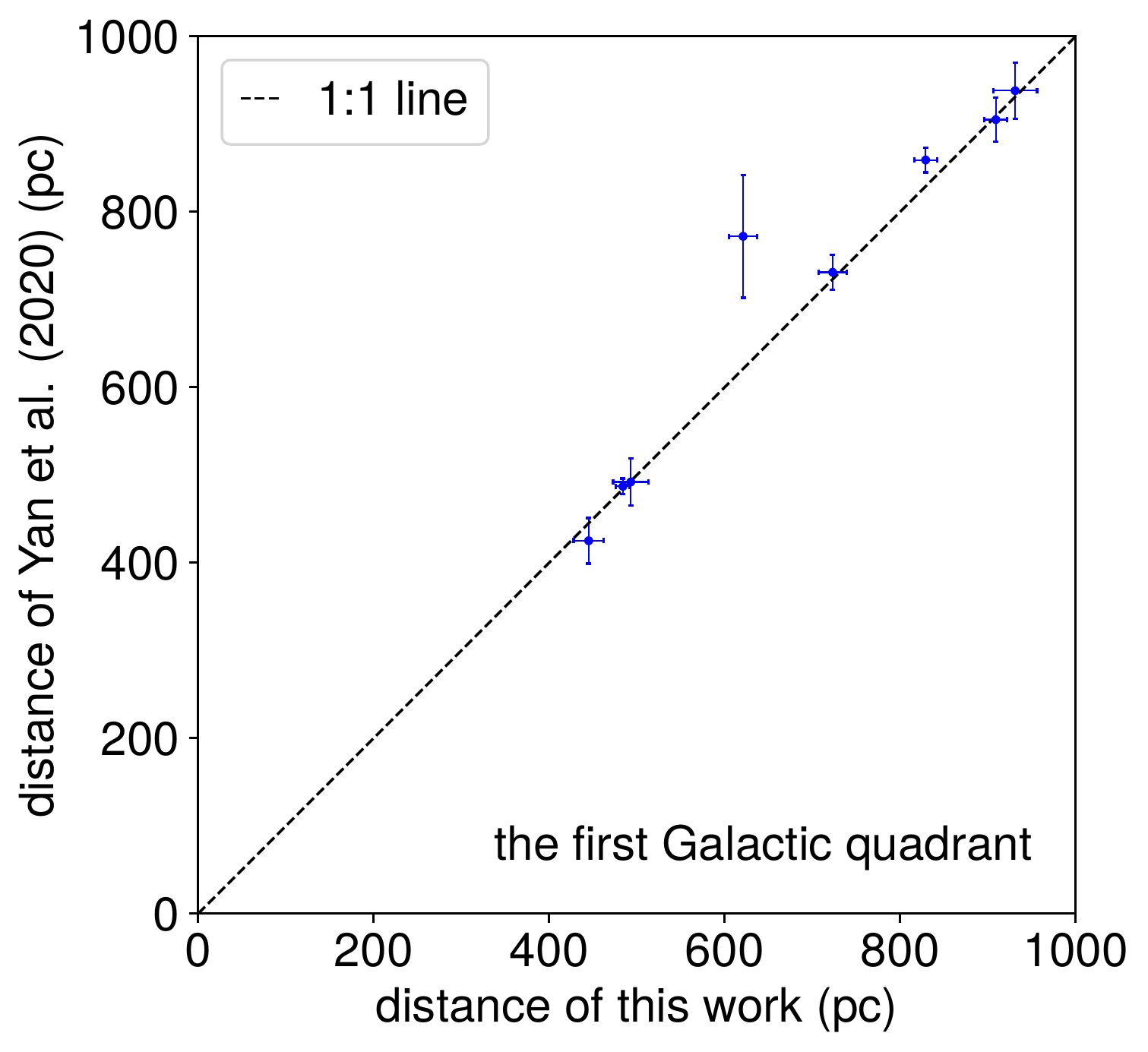}{0.32\textwidth}{(a)}    \fig{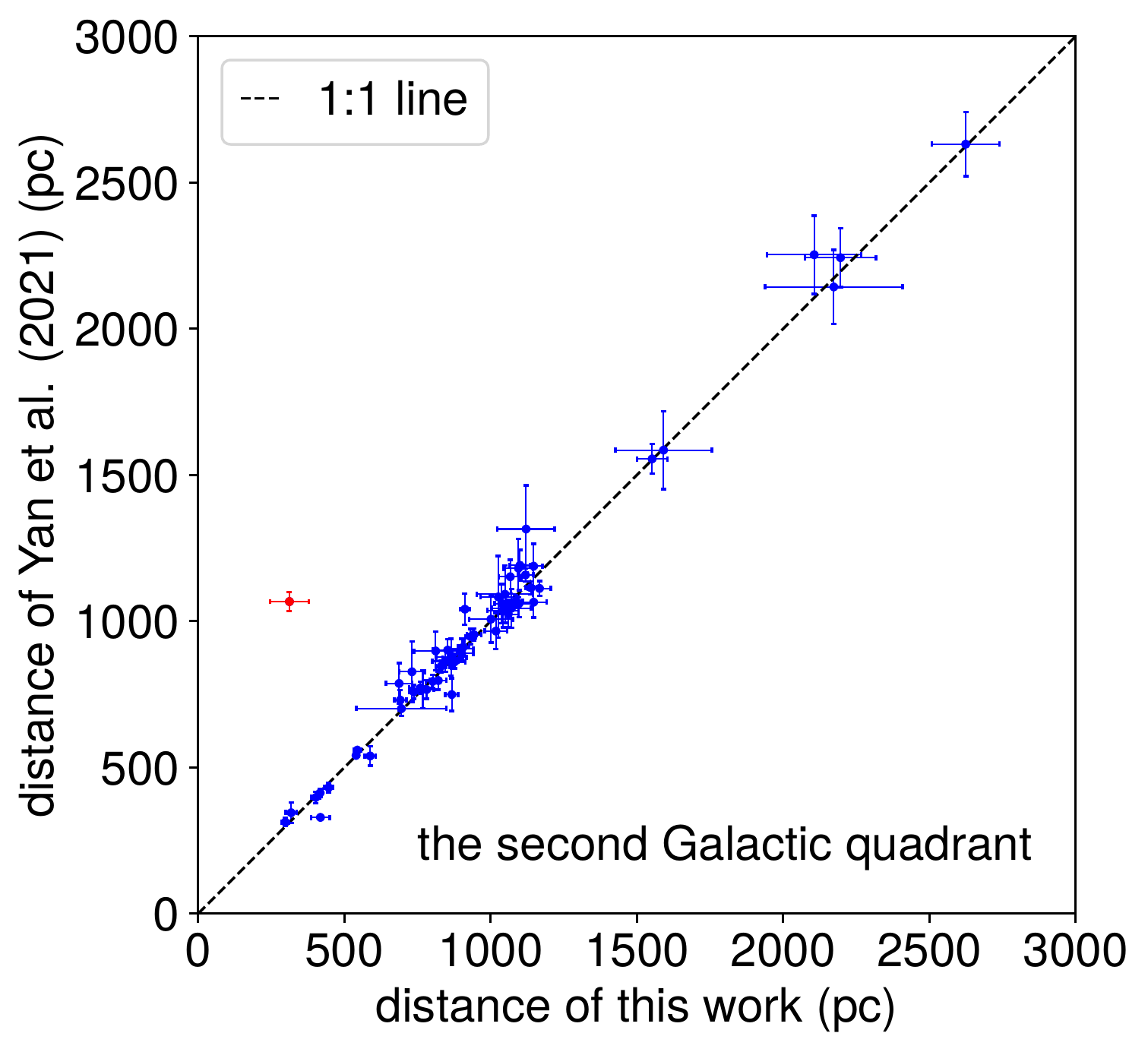}{0.32\textwidth}{(b)}   \fig{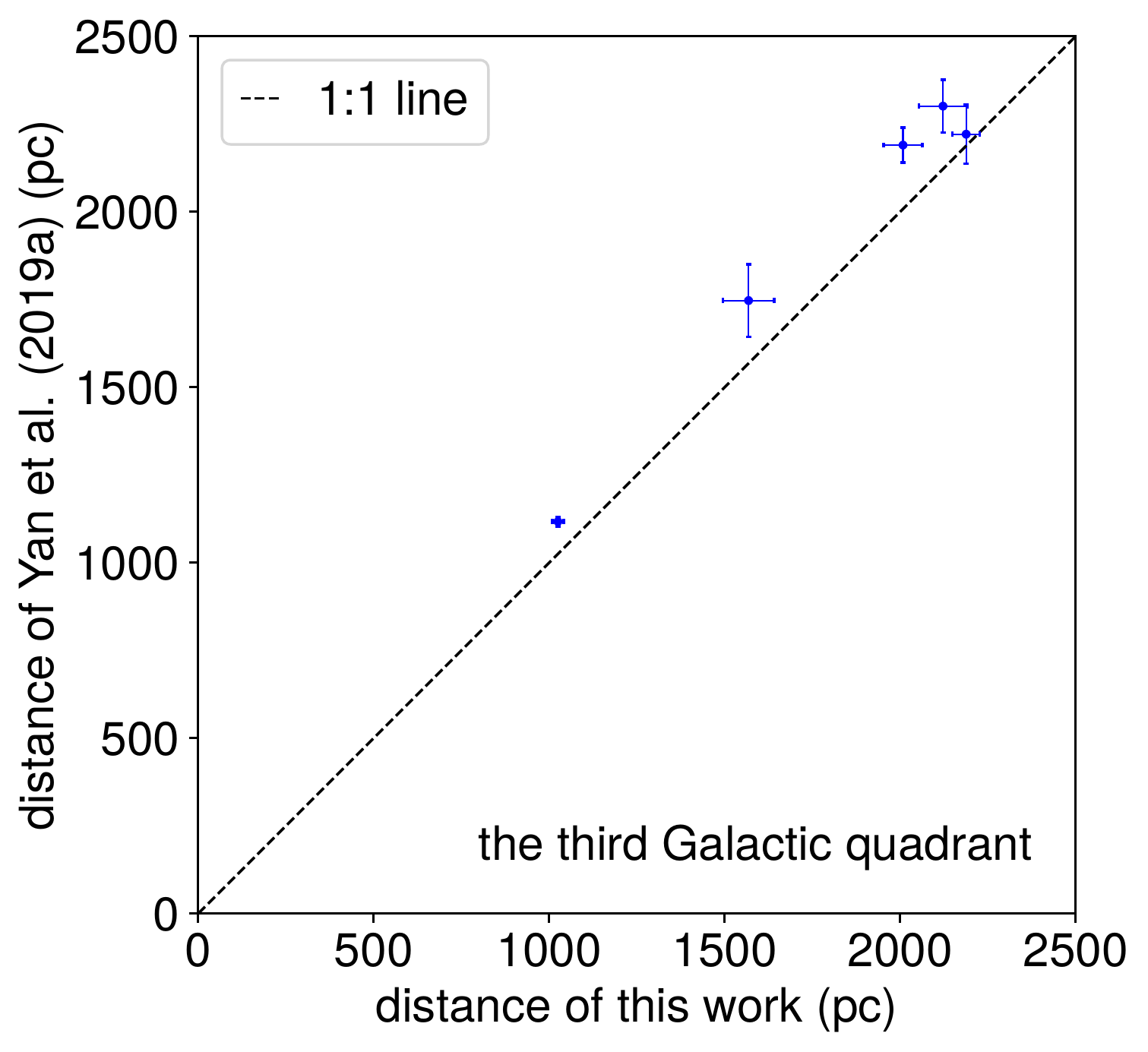}{0.32\textwidth}{(c)} 
 } 
 
\caption{Comparisons of distances with  previous BEEP results: (a) eight molecular clouds in the first Galactic quadrant \citep{2020ApJ...898...80Y}, (b) 63 molecular clouds in the second Galactic quadrant \citep{2021A&A...645A.129Y}, and (c) five molecular clouds in the third Galactic quadrant  \citep{2019ApJ...885...19Y}. One molecular cloud, G148.1$-$00.2, marked in red, shows a large distance discrepancy. \label{fig:compareq2} }
\end{figure}




 \section{Discussion}
   \label{sec:discuss}

\subsection{Foreground Ambiguity}

To demonstrate the effect of foreground contamination, we display distances to G123.2$-$00.5 in Figure \ref{fig:foreground}. The foreground distance of G123.2$-$00.5 is about 1000 pc, close to that the distance to G123.2$-$00.5 itself. In this case, it is unclear which distance is measured, so G123.2$-$00.5 is rejected.

In total, 14 molecular clouds are rejected due to the foreground ambiguity. Eleven are in the second Galactic quadrant, and three are in the third Galactic quadrant. In the second Galactic quadrant, according to the radial velocity \citep{2020ApJS..246....7S}, six of the eleven molecular clouds are located in the Perseus arm, and the rest five are in the local arm.

\subsection{Comparison With Previous Studies}
In this section, we compare molecular cloud distances with previous BEEP results \citep{2019ApJ...885...19Y,2020ApJ...898...80Y,2021A&A...645A.129Y}. \citet{2021A&A...645A.129Y} use the same molecular cloud catalog (except for the including of incomplete clouds), so we first compare the results in the second Galactic quadrant.

\begin{figure}[ht!]
 \plotone{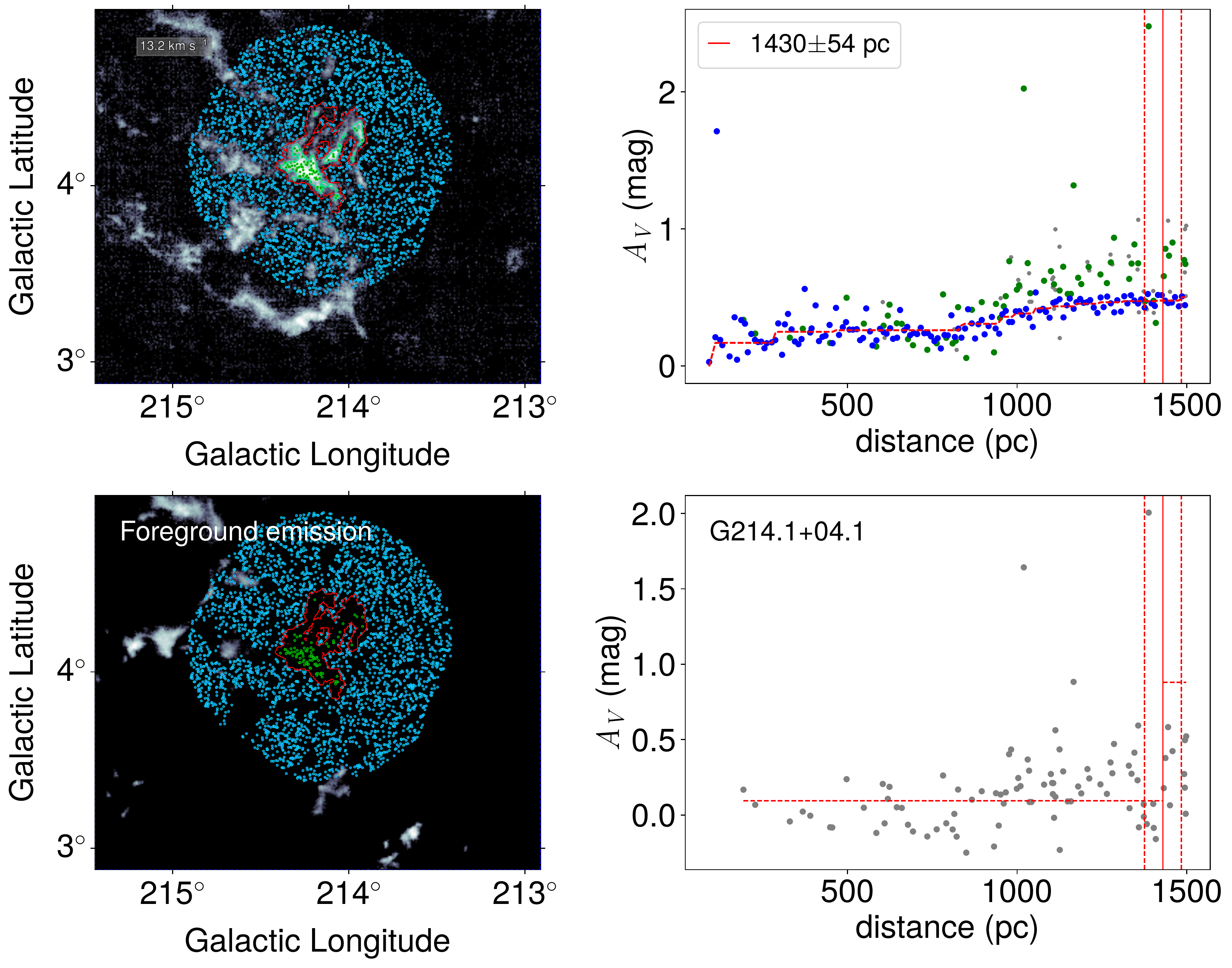}   

\caption{Same as Figure \ref{fig:mcmc} but for a molecular cloud whose distance is not measured with MCMC sampling. The distance given by piecewise linear regression in  the BEEP-II method is about 950 pc, and the non-detection of MCMC sampling is due to several low extinction background stars. See Figure \ref{fig:mcmc} for other details. \label{fig:failmcmc} } 
\end{figure}

\citet{2021A&A...645A.129Y} measured distances to 77 molecular clouds (including the largest molecular cloud), and in this work, we measured distances to 155 molecular clouds, of which 64 are in common. Among the commonly detected molecular clouds, 63 have unique distances, which are plotted in panel (b) of Figure \ref{fig:compareq2}. Most molecular cloud distances are compatible within errors, and one molecular cloud, G148.1$-$00.2, shows a large discrepancy, i.e., $>$  25\% of this work. The extinction jump of G148.1$-$00.2 is small compared with that caused by background molecular clouds, implying that the measurement of the distance of background molecular clouds made by \citet{2021A&A...645A.129Y} is incorrect.

 The BEEP-II method slightly improves the precision of distances. For 63 molecular clouds in the second Galactic quadrant shown in Figure \ref{fig:compareq2}, the mean relative statistic error of the BEEP-II distances is 4.81\%,  about 0.04\% smaller than those of the BEEP distances.



 In the results of \citet{2021A&A...645A.129Y}, nine molecular clouds are excluded due to the incompleteness in PPV space, and distances of four molecular clouds are not measured in this work. It is interesting to examine the reason for non-detections, which potentially reveals shortages of the BEEP-II method. Two molecular clouds, G112.2$-$01.5 and G131.8$+$00.0, have their distances measured in the clustering step, and both distances are close to that of \citet{2021A&A...645A.129Y}. However, their distance errors provided by MCMC sampling are large. The rest two molecular clouds, G114.5$-$00.1 and G131.8$-$02.5, did not yield distance clusters  due to the unclear extinction jump, and more accurate extinction data are needed to confirm their distances.


 
 

  

In the first Galactic quadrant, \citet{2020ApJ...898...80Y} use SCIMES to split large structures in PPV space, and the molecular cloud samples are largely different. However, some small molecular clouds were not split by SCIMES, and can be matched with DBSCAN molecular cloud samples, where the matching criteria are that both $l$ and $b$ differences are less than 0\fdg1 and the $V_{\rm LSR}$ difference is less than  0.5 \kms.  As shown in panel (a) of Figure \ref{fig:compareq2}, eight molecular clouds have their distances commonly measured by both this work and \citet{2020ApJ...898...80Y}, and their results agree well with each other. 


In the third Galactic quadrant, we measured many more molecular cloud distances than \citet{2019ApJ...885...19Y}. This is because \citet{2019ApJ...885...19Y} only measured large molecular clouds, ignoring small ones, and in addition, they only used \ag\ data. Using the same match criteria mentioned above, we found distances of five molecular clouds  are commonly measured. The distances of \citet{2019ApJ...885...19Y} are systematically larger (by about 133 pc) than that of this work, as demonstrated in panel (c) of Figure \ref{fig:compareq2}. In this work, distances of those five molecular clouds are measured with \av, while they are measured with \ag\ in  \citet{2019ApJ...885...19Y}, so this systematic discrepancy is mostly due to the \ag\ and \av\ data. 


   

   

\subsection{Structure of DBSCAN Clouds}

 The global search feature of the BEEP-II method makes it capable of examining the structure of molecular clouds identified with DBSCAN. Limited by the precision of stellar parallaxes and extinction, however, detailed analyses of 3D structures of molecular clouds are beyond the scope of this work, but the BEEP-II method can roughly provide the number of distance components for each molecular cloud.

Among the 234 molecular clouds, one, G134.7$-$00.3, shows multiple distance components. In terms of the angular area, G134.7$-$00.3 is the second largest molecular cloud with an angular area of 11 deg$^2$, while the next large molecular cloud has an area of about 7 deg$^2$. Roughly, G134.7$-$00.3 has three components at about 740, 900, and 1400 pc. \citet{2021A&A...645A.129Y} also derived 22 distances for a large-scale molecular cloud, which is excluded in this work due to its incompleteness.  Consequently, for small molecular clouds ($<$ 7 deg$^2$), the velocity crowding effect in PPV space is not severe,  and at least local molecular clouds are not entangled with each other. However, there is still a chance that local molecular clouds overlap with distant components ($>$ 3 kpc) in PPV space, but distinguishing them is beyond the scope of Gaia DR2 data.

\subsection{Ability of the BEEP-II Method}

The BEEP-II method examines the distance twice, first by piecewise linear regression and then by MCMC sampling. There are molecular clouds whose distances are detected with piecewise linear regression but not with MCMC sampling, e.g., see Figure \ref{fig:failmcmc} for the MCMC sampling result of G214.1$+$04.1. The failure of MCMC sampling is due to the unclear extinction jump. The distance given by piecewise linear regression, however, is about 950 pc, consistent with the jump position as judged by eye.

The BEEP-II method is a powerful tool for measuring molecular cloud distances. First, the automatic selection of on- and off-cloud star sample pairs improves the efficiency of the BEEP method significantly. The computational time (on an intel Core i9-10920X desktop in Ubuntu operating system and in \texttt{Python} code) used to examine molecular cloud distances in the first, second, and third Galactic quadrants is about 2 (359 clouds), 10 (1618 clouds), and 2 hours (237 clouds), respectively.  At present, we confine the distance search within 3 kpc, but updated Gaia and extinction data in future will enable distance measurements for more distant molecular clouds. Secondly, the BEEP-II method determines the optimal region for distance measuring. These two features are the essential improvements of the BEEP-II method over the BEEP method.

 The number of molecular cloud distances measured by the BEEP-II method has surpassed that of masers. \citet{2019ApJ...885..131R} reported 199 maser parallaxes measured by the Bar and Spiral Structure Legacy (BeSSeL) Survey and the Japanese VLBI Exploration of Radio Astrometry  (VERA) project, while the BEEP-II method measures distances to 234 molecular clouds.

\subsection{Molecular Cloud Catalogs}

 Large-scale CO surveys are gradually transforming molecular cloud researches from case studies to sample studies. For instance, using the CfA-Chile survey,  \citet{2016ApJ...822...52R} and \citep{2017ApJ...834...57M} built two different molecular cloud catalogs with different algorithms. Based on the $^{13}\mathrm{CO}~(J=2\rightarrow1)$ data of the  Structure, Excitation and Dynamics of the Inner Galactic Interstellar Medium \citep[SEDIGISM, ][]{2021MNRAS.500.3064S} survey and the SCIMES algorithm, \citet{2021MNRAS.500.3027D} find 10663 molecular clouds. With the same algorithm but using the JCMT  $^{12}\mathrm{CO}~(J=3\rightarrow2)$ High-Resolution Survey \citep[COHRS,][]{2013ApJS..209....8D} data, \citet{2019MNRAS.483.4291C} build a catalog with more than 85000 molecular clouds. 
 
 Statistical studies of molecular cloud properties hinge on the distance information, but the fraction of molecular clouds with accurate distances is small. In this work, we collectively identified 2214 molecular clouds  with angular areas larger than 0.015 deg$^2$ and complete in PPV space, but the fraction of samples with measured distances is only  11\%. In addition to physical property statistics, delineating the large-scale distribution of local molecular clouds also needs high completeness of molecular cloud distances. Consequently, distance measurements are still far from keeping up with demand.






\section{Summary}
\label{sec:summary}

We have improved the BEEP method and developed a procedure of measuring distances to molecular clouds through a global search method, which is called the BEEP-II method. This improved method examines distances across the surface of molecular clouds, analyzing distance components accordingly. Compared with the BEEP method, the BEEP-II method is suitable for measuring distances of molecular clouds with large angular size or in complicated environments. In addition, the high efficiency of the BEEP-II method enables systematic distance measurements of molecular clouds mapped by large-scale CO surveys.




In three regions mapped by the MWISP CO survey, the BEEP-II method measures distances to 234 molecular clouds in total, including 238 positions, more than double the number of previous BEEP results. The statistical error is about 7.28\% on average, and comparisons of distances in the second Galactic quadrant indicate that the BEEP-II method improves the distance precision by about 0.04\%. The augmentation of molecular cloud samples with high-precision distances will help reveal the distribution and properties of molecular clouds.







\begin{acknowledgments}
We thank Sammy McSweeney for his careful proofreading. We are grateful to all the members of the MWISP working group, particularly the staff members at PMO-13.7m telescope, for their long-term support. 
MWISP was sponsored by National Key R\&D Program of China with grant no. 2017YFA0402701 and  CAS Key Research Program of Frontier Sciences with grant no. QYZDJ-SSW-SLH047. JY is supported by National Natural Science Foundation of China through grant 12041305. This work was also sponsored by National Natural Science Foundation of China grant No. 12003071, 11773077, U1831136,  and the Youth Innovation Promotion Association, CAS (2018355).  
\end{acknowledgments}

%

\vspace{5mm}
\facilities{PMO 13.7 m, GAIA}


\software{astropy \citep{2013A&A...558A..33A},     
						SciPy \citep{2020NatMe..17..261V}
          }

 \bibliographystyle{aasjournal}
 \bibliography{refGAIADIS}





%

\end{document}